\documentclass[final,prd,amsfonts,amsmath,amssymb,nobibnotes,nofootinbib,eqsecnum,showpacs,showkeys]{revtex4}
\usepackage{type1cm,bm,dcolumn}
\usepackage[dvips]{graphicx,color}
\newcommand{\be}{\begin{equation}}
\newcommand{\ee}{\end{equation}}
\newcommand{\bea}{\begin{eqnarray}}
\newcommand{\eea}{\end{eqnarray}}
\newcommand{\ben}{\begin{enumerate}}
\newcommand{\een}{\end{enumerate}}
\newcommand{\bi}{\begin{itemize}}
\newcommand{\ei}{\end{itemize}}

\newcommand{\Ord}[2]{\mathcal O \left(#1\right)^{#2}}

\begin{document}

%:Title
\title{Inflating baby-Skyrme branes in six dimensions}
\author{Yves Brihaye}
\email{yves.brihaye(at)umons.ac.be}
\author{T\'{e}rence Delsate}
\email{terence.delsate(at)umons.ac.be}
\affiliation{Theoretical and Mathematical Physics Dpt.,\\
Universit\'{e} de Mons - UMons, 20, Place du Parc, 7000 Mons - Belgium}
\author{Yuta Kodama}
\email{yutakodama(at)gmail.com}
\author{Nobuyuki Sawado}
\email{sawado(at)ph.noda.tus.ac.jp}
\affiliation{Department of Physics, Faculty of Science and Technology,
Tokyo University of Science, Noda, Chiba 278-8510, Japan}
\date{\today}

%:Abstract
\begin{abstract}
We consider a six dimensional brane world model, where the brane is described by a localized solution to the baby-Skyrme model extending in the extradimensions. The branes have a cosmological constant modeled by inflating four dimensional slices and we further consider a bulk cosmological constant. We construct solutions numerically and present evidences that the solutions cease to exist for large values of the brane cosmological constant in some particular case.
Then we study the stability of the model by considering perturbation of the gravitational part (resp. baby skyrmion) with fixed matter fields (resp. gravitational background). Our results indicate that the perturbation equations do not admit localized solutions for certain type of perturbation. The stability analysis can be alternatively seen as leading to a particle spectrum; we give mass estimations for the baby-Skyrme perturbation and for the graviton.
\end{abstract}

%:Pacs & Keywords
\maketitle

\section{Introduction}
Theories with extradimensions have been expected to solve the hierarchy 
problem and cosmological constant problem. Experimentally unobserved 
extradimensions indicate that the standard model particles and forces 
are confined to a 3-brane~\cite{ArkaniHamed:1998rs,ArkaniHamed:1998nn,Randall:1999ee,Randall:1999vf}. 
Intensive study has been performed for the Randall-Sundram (RS) brane model in 5 space-time 
dimensions~\cite{Randall:1999ee,Randall:1999vf}. In this framework, the exponential 
warp factor in the metric can generate a large hierarchy of scales. 
This model, however, requires unstable negative tension branes  
and the fine-tuning between brane tensions and bulk cosmological constant.
 
There is hope that higher dimensional brane models more than five dimensions
could evade those problems appeared in 5-dimensions. 
In fact brane theories in 6-dimensions show a very distinct feature towards 
the fine-tuning and negative tension brane problems. 
Warped compactifications are possible in 6 space-time dimensions 
in the model of topological objects such as defects and solitons.  
In this context abelian strings~\cite{Cohen:1999ia,Gregory:1999gv,Gherghetta:2000qi,
Giovannini:2001hh,Peter:2003zg} 
were investigated, showing that they can realize localization of gravity for negative
cosmological constant. (For the positive case, solutions which localise gravity are 
lost~\cite{Brihaye:2003ur}.)  
For the magnetic monopoles, similar compactification was achieved for both positive 
and negative cosmological constant~\cite{Roessl:2002rv}. 
Interestingly, if the brane is modeled in such a field theory language, 
the fine-tuning between bulk and brane parameters required in the case 
of delta-like branes turns to a tuning of the model parameters~\cite{Peter:2003zg}. 

It is well known that there are two main contexts in which solitons appear in field theories. 
One is like the strings and the magnetic monopoles inabelian and non-abelian gauge theories, and the others are
kinds of non-linear type models, such as the skyrmions, hopfions~\cite{Skyrme:1961vq,Faddeev:1996zj}. 
The latter are particularly interesting and have deep insight in their non-trivial topological structures. 
The Skyrme model is known to possess soliton solutions called baby skyrmions 
in 2-dimensional space~\cite{Piette:1994jt,Piette:1994ug,Kudryavtsev:1996er}. 
The warped compactification of the 2-dimensional 
extra space by such baby skyrmions has already been studied~\cite{Kodama:2008xm}.  
The authors found regular, static solutions with non-trivial topology 
which realize warped compactification for a negative bulk cosmological constant. 
Also, a somewhat different model, namely Maxwell gauged $\mathbb{C}P^{1}$ 
type of non-linear $\sigma$ model has been investigated in \cite{Kodama:2007kr}.
 
Many of previous studies were based on the assumption that the cosmological constant 
inside the 3-branes is tentatively set to be zero; the branes were assumed to be static Minkowski, 
``flat 3-branes'', 
despite the fact that our universe has a small but positive definite cosmological constant. 
Thus, addressing the non-zero cosmological constant inside the branes has been considered first by Cho and Vilenkin in \cite{chovil} and then extended for case of the strings~\cite{Brihaye:2006pi} and 
the monopoles~\cite{Brihaye:2006cs}. 
They have studied for both signs of the bulk cosmological constant. 
In this paper, we introduce ``inflation'' on the baby-skyrmion branes with both signs of 
bulk cosmological constant. 

Another aim of the present paper is to analyze linear stability of our new solutions by 
fluctuating all fields.
Analysis for gravitating thick defects embedded in higher dimensions are found in the literature;
for 5-dimensions {\cite{Giovannini:2001xg,Giovannini:2002jfa,Giovannini:2006ye}}, and 
for 6-dimensions {\cite{Giovannini:2002mk,Giovannini:2002sb,Peter:2003zg,RandjbarDaemi:2002pq}}.
(Note that the model in \cite{Giovannini:2006ye} is constructed by gravitating multidefects 
in five dimensions).
The studies for thick defects, however, are works in progress
since the topological defects used in the literature are complicated structures.
In this paper, we introduce a general form of perturbation of the metric and the skyrmion fields. 
Employing a simplified ansatz and gauge conditions, we find that two independent equations are 
sufficient to discuss the stablity. We shall solve two of them independently and present 
typical localized eigenstates of the metric and the matter field fluctuations. 
We shall finally give some speculations for physics of our universe, i.e., 
SM particle spectra, a signature of the CMB and others. 

This paper is organized as follows. In the next section we describe the 
Einstein-Skyrme system in 6-dimensions and derive coupled equations for 
the Skyrme and gravitational fields. Several types of solutions are found in Sec.III. 
Sec IV is the analysis of the asymptotic behavior of the solutions at the infinity. 
In Sec V, we present our numerical results. 
We then give a detailed analysis for the gravity and the matter perturbations in Sec.VI. 
Conclusion and discussion are given in Sec.VII.

\section{The gravitating baby-Skyrme Model in six dimensions}

The total action for the gravitating baby-Skyrme model is of the form $S = S_{\text{grav}} + S_{\text{baby}}$.
The gravitational part
\begin{equation}
S_{\text{grav}} = \int d^{6}x\sqrt{-g}\left( \frac{1}{2\chi_{(6)}}R - \Lambda_{(6)} \right), 
\label{eq:gravity-action}
\end{equation}
is the generalized Einstein-Hilbert gravity action, where $\Lambda_{(6)}$ is the bulk cosmological constant, and $\chi_{(6)} = 8\pi G_{(6)} = 8\pi/M_{(6)}^{4}$.

On the other hand, the action for the baby-Skyrme model $S_{\text{baby}}$ is given by
\be
S_{\text{baby}} = \int d^{6}x\sqrt{-g}\Bigl[\frac{\kappa_{2}}{2}(\partial_{M}\bm{n})\cdot(\partial^{M}\bm{n})-\frac{\kappa_{4}}{4}(\partial_{M}\bm{n}\times\partial_{N}\bm{n})^{2}- \kappa_{0}V(\bm{n}) \Bigr].
\label{eq:brane-action}
\ee
We will use the convenient notation $S_{\text{baby}} \equiv \int d^{6}x\sqrt{-g}\mathcal{L}_{\text{baby}}$.
Here $\bm{n}$ is a scalar triplet subject to the nonlinear constraint $\bm{n}\cdot\bm{n}=1$, and $V(\bm{n})$ is the potential term with no derivatives of $\bm{n}$.
The coefficients $\kappa_{2,4,0}$ in Eq.(\ref{eq:brane-action}) are the coupling constants in the gravitating baby-Skyrme model.

\subsection{The Ansatz}
Assuming axial symmetry for the extradimensions, the metric can be written in the following form
\begin{equation}
	ds^{2} = M^{2}(\rho)g_{\mu\nu}^{(4)}dx^{\mu}dx^{\nu} - d\rho^{2} - l^{2}(\rho)d\theta^{2}
	\label{eq:6D-metric}
\end{equation}
where $\rho \in [0, \infty)$ and $\theta \in [ 0, 2\pi]$ are the coordinates associated with the extra dimensions.

We further model a cosmological constant on the brane by considering the following form of the four dimensional 
subspace (described by $g_{\mu\nu}^{(4)}$ in Eq.\eqref{eq:6D-metric})
\begin{equation}
	ds_{(4)}^{2} = g_{\mu\nu}^{(4)}dx^{\mu}dx^{\nu} = dt^{2} - \delta_{ij}e^{2H(t)}dx^{i}dx^{j}
	\label{eq:4D-metric}
\end{equation}
where $H(t)$ is a function of the time coordinate $t$.

The ansatz for the scalar triplet $\bm{n}$ is given by the hedgehog Ansatz \cite{Piette:1994jt}:
\begin{equation}
	\bm{n} = (\sin f(\rho)\cos(n\,\theta), \sin f(\rho)\sin(n\,\theta), \cos f(\rho)).
	\label{eq:hedgehog-ansatz}
\end{equation}

Let us note that there are some variations {\cite{Eslami:2000tj}} for choosing the potential term $V(\bm{n})$
in the baby-Skyrme model (\ref{eq:brane-action}).
Here we use the so-called \emph{old} baby skyrmions potential, which reads
\begin{equation}
	V(\bm{n}) = 1 - \bm{n}\cdot\bm{n}^{(\infty)} = 1 + \cos f(\rho),
	\label{eq:potential}
\end{equation}
where $\bm{n}^{(\infty)} = \lim_{\rho\to\infty}\bm{n}(\rho,\theta)$ is the vacuum configuration of the baby-Skyrme model.

\subsection{Field equations of the model}
In order to rewrite the system in terms of dimensionless quantities, we define
\begin{equation}
	r := \sqrt{\frac{\kappa_{2}}{\kappa_{4}}}\,\rho,\;\;\;
	L(r) := \sqrt{\frac{\kappa_{2}}{\kappa_{4}}}\,l(r),\;\;\;
	\mathcal{H}(t) := \sqrt{\frac{\kappa_{4}}{\kappa_{2}}}H(t).
	\label{eq:dimensionless}
\end{equation}

We also introduce dimensionless parameters
\footnote{The dimensions of the model parameters are:
$[\Lambda_{(6)}] = M^{6}$, $[\chi_{(6)}] = M^{-4}$,
$[\kappa_{2}] = M^{4}$, $[\kappa_{4}] = M^{2}$ $[\kappa_{0}] = M^{6}$.} according to
\be
\alpha := \chi_{(6)}\kappa_{2},\ \beta := \Lambda_{(6)}\kappa_{4}/\kappa_{2}^{2},\ \mu :=\kappa_{0}\kappa_{4}/\kappa_{2}^{2}.
\ee

%$\alpha := \chi_{(6)}\kappa_{2}$ is the dimensionless gravitational coupling constant,
%$\beta := \Lambda_{(6)}\kappa_{4}/\kappa_{2}^{2}$ is the dimensionless bulk cosmological constant,
%and $\mu := \kappa_{0}\kappa_{4}/\kappa_{2}^{2}$ is the dimensionless strength of the potential term $V(\bm{n})$%

Finally, we introduce
\be 
u(r) := 1 + \frac{n^{2}}{L^{2}(r)}\sin^{2}f(r),\ v(r) := 1 - \frac{n^{2}}{L^{2}(r)}\sin^{2}f(r)
\nonumber\ee
for latter convenience and use the notation $f' := \partial_{r}f(r)$.

After short computations, the baby Skyrme equations and Einstein equations reduce to
\be
	u\,f'' + \left( 4\frac{M'}{M} + \frac{L'}{L} + \frac{u'}{u} \right)u\,f'
		- (1 + f'^{2})\frac{n^{2}}{L^{2}}\sin f\cos f + \mu\sin f = 0,
	\label{eq:eq-baby}
\ee
for the baby Skyrme field and
\be
	3\frac{M'^{2}}{M^{2}} + \frac{L''}{L} + 3\frac{M'L'}{ML} + 3\frac{M''}{M}
		- 3\frac{\mathcal{H}_{,t}^{2}}{M^{2}} = \alpha( \tau_{0} - \beta ),
		\label{eq:Einstein-eq_t}
\ee
\be
3\frac{M'^{2}}{M^{2}} + \frac{L''}{L} + 3\frac{M'L'}{ML} + 3\frac{M''}{M} - \frac{2\mathcal{H}_{,t,t}+3\mathcal{H}_{,t}^{2}}{M^{2}} = \alpha( \tau_{0} - \beta ),
\label{eq:Einstein-eq_3}
\ee
\be
6\frac{M'^{2}}{M^{2}} + 4\frac{M'L'}{ML} - 3\frac{\mathcal{H}_{,t,t}+2\mathcal{H}_{,t}^{2}}{M^{2}}= \alpha( \tau_{r} - \beta ),
\label{eq:Einstein-eq_r}
\ee
\be
4\frac{M''}{M} + 6\frac{M'^{2}}{M^{2}} - 3\frac{\mathcal{H}_{,t,t}+2\mathcal{H}_{,t}^{2}}{M^{2}}= \alpha( \tau_{\theta} - \beta ),
\label{eq:Einstein-eq_th}
\ee
for the Einstein equation. We used the notation $\mathcal{H}_{,t} := \partial_{t}\mathcal{H}(t)$ and $\mathcal{H}_{,t,t} := \partial_{t}^{2}\mathcal{H}(t)$.

Let us note that Eq.(\ref{eq:Einstein-eq_t}) and (\ref{eq:Einstein-eq_3}) are the four-dimensional components of the Einstein equation, while Eq.(\ref{eq:Einstein-eq_r}) and (\ref{eq:Einstein-eq_th}) are the extra-dimensional components.

The components of the dimensionless energy-momentum (EM) tensor in Eqs.(\ref{eq:Einstein-eq_t})-(\ref{eq:Einstein-eq_th}) are given by
\begin{equation}
\begin{aligned}
	&\tau_{0} = - \frac{1}{2}u\,f'^{2} - \frac{1}{2}\frac{n^{2}}{L^{2}}\sin^{2}f - \mu(1 + \cos f),\\
	&\tau_{r} = \frac{1}{2}u\,f'^{2} - \frac{1}{2}\frac{n^{2}}{L^{2}}\sin^{2}f - \mu(1 + \cos f),\\
	&\tau_{\theta} = - \frac{1}{2}v\,f'^{2} + \frac{1}{2}\frac{n^{2}}{L^{2}}\sin^{2}f - \mu(1 + \cos f),
\end{aligned}
\label{stresstensor}
\end{equation}

We supplement the system of equations \eqref{eq:eq-baby}-\eqref{eq:Einstein-eq_th} by the following set of boundary conditions:
\begin{equation}
	f(0) = -(m-1)\pi,\;\;\; f(\infty) = \pi,
	\label{eq:BC-f}
\end{equation}
where $m, n \in \mathbb{Z}$, for the baby Skyrme field and
\be
L(0)=0,\ L'(0)=1,\ M(0)=1,\ M'(0)=0,
\label{eq:BC-gr}
\ee
for the metric fields.

The above boundary conditions are required for regularity and finiteness of the energy.

Note that considering the hedgehog Ansatz (\ref{eq:hedgehog-ansatz}) under 
the boundary condition (\ref{eq:BC-f}), one can construct a topological charge (or winding number) defined as follows
\be
N = \frac{1}{4\pi}\int \bm{n}\cdot(\partial_{\rho}\bm{n}\times\partial_{\theta}\bm{n}) d\rho\,d\theta= \frac{n}{2}\left[ 1 + (-1)^{1-m} \right] \in \mathbb{Z}.
\label{eq:winding}
\ee

\subsection{Inflating four-dimensional slices}
A solution for $\mathcal{H}(t)$ is actually trivial. Combining Eq.\eqref{eq:Einstein-eq_3} and Eq.\eqref{eq:Einstein-eq_t},
leads to $\mathcal{H}_{,t,t} = 0$, \emph{i.e.} $\mathcal{H}_{,t} = \text{const} \equiv \mathcal{H}_{0}$.
Thus we obtain the simple solution given by
\begin{equation}
	\mathcal{H}(t) = \mathcal{H}_{0}\,t
	\label{eq:solution-H}
\end{equation}
where $\mathcal{H}_{0}$ is a constant; and where we neglected an integration constant. Note that the integration constant can be absorbed by a global rescaling of the three dimensional spacial coordinates.

The result (\ref{eq:solution-H}) follows from the Ansatz for the four-dimensional metric (\ref{eq:4D-metric})
and the important assumption under which the gravitational source $\mathcal{L}_{\text{baby}}$ is a function of the extra-dimensional coordinates only.

Therefore we introduce a new dimensionless parameter
\begin{equation}
	\gamma := \mathcal{H}_{0}^{2} \equiv \frac{\kappa_{4}}{\kappa_{2}}H_{0}^{2},
\end{equation}
and we arrange the components of the Einstein equation in terms of $\gamma$:
\bea
&&3\frac{M'^{2}}{M^{2}} + \frac{L''}{L} + 3\frac{M'L'}{ML} + 3\frac{M''}{M}- 3\frac{\gamma}{M^{2}} 
= \alpha( \tau_{0} - \beta ),
\label{eq:Einstein-eqs1}\\
&&6\frac{M'^{2}}{M^{2}} + 4\frac{M'L'}{ML} - 6\frac{\gamma}{M^{2}}= \alpha( \tau_{r} - \beta ),
\label{eq:Einstein-eqs2}\\
&&4\frac{M''}{M} + 6\frac{M'^{2}}{M^{2}} - 6\frac{\gamma^{2}}{M^{2}}= \alpha( \tau_{\theta} - \beta ).
\label{eq:Einstein-eqs3}
\eea

Note that it is possible to interpret of $\gamma$ using the four dimensional effective theory following 
the lines of ref. \cite{Brihaye:2006pi}.

Seen differently, $\gamma$ can be interpreted as a positive cosmological constant in the 
four dimensional subspace of the full model, since in this case, $g^{(4)}_{ab}$ is such 
that $G^{(4)}_{ab} = 3H^2 g^{(4)}_{ab}$, where $G^{(4)}_{ab}$ is the Einstein tensor computed with $g^{(4)}_{ab}$. 
Note that replacing the four dimensional subspace by another Einstein spacetime satisfying 
$G^{(4)}_{ab} = 3H^2 g^{(4)}_{ab}$, such as the Schwarzschild-de Sitter spacetime, leads to the same equations.

Another useful quantity is the rescaled Ricci scalar which will be used later and is given by
\be
R = \frac{2 L''}{L}+\frac{8 L' M'}{L M}+\frac{8M''}{M}+\frac{12 (M')^2}{M^2}-\frac{12 \gamma }{M^2}.
\ee

\section{Special solutions for topological vacuum}
Let us for a moment consider the model with $f(r) \equiv \pi$, i.e. where the baby-Skyrme field is a topological vacuum configuration.

It is known that the Einstein equation (\ref{eq:Einstein-eqs1})-(\ref{eq:Einstein-eqs3}) for the topological vacuum have several special solutions.
Such solutions are classified according to the model parameters, mainly $\beta$ and $\gamma$.
Before reviewing these special vacuum solutions, we shall present a useful relation derived in \cite{Brihaye:2006pi}.
From Eq.(\ref{eq:Einstein-eqs2}),(\ref{eq:Einstein-eqs3}), we have
\begin{equation}
	\frac{M''}{M} = \frac{M'L'}{ML} \iff L(r) = C_{L}\frac{dM(r)}{dr}
	\label{eq:useful-relation}
\end{equation}
where $C_{L}$ is an integration constant.
The relation (\ref{eq:useful-relation}) is quite general except in some special cases, \emph{e.g.} for $M = 0$ and $M' = 0$.

\subsubsection{Flat branes case: $\gamma = 0$}
Special solutions for $\gamma = 0$ have been investigated by many authors in Refs. {\cite{Giovannini:2001hh,Brihaye:2003ur,Brihaye:2006cs}}. We shortly summarize them here.
Let $C_{1}$, $C_{2}$ and $r_{0}$ be constants of integrations; for $\beta = 0$, Eqs.(\ref{eq:Einstein-eqs1})-(\ref{eq:Einstein-eqs3}) admit two different types of solutions: the string branch
\begin{equation}
	M_{s}^{0}(r) = C_{1},\;\;\; L_{s}^{0}(r) = C_{2}(r - r_{0}),
\end{equation}
and the Melvin branch
\begin{equation}
\begin{aligned}
	&M_{m}^{0}(r) = C_{1}(r - r_{0})^{2/5},\\
	&L_{m}^{0}(r) = C_{2}(r - r_{0})^{-3/5}.
\end{aligned}
\end{equation}
For $\beta > 0$, Eqs.(\ref{eq:Einstein-eqs1})-(\ref{eq:Einstein-eqs3})
 have a set of the periodic solutions given by
\begin{equation}
\begin{aligned}
	&M_{p}^{0}(r) = C_{1}\cos^{2/5}\sqrt{\frac{5\alpha\beta}{8}}(r-r_{0}),\\
	&L_{p}^{0}(r) = C_{2}\frac{\sin\sqrt{\frac{5\alpha\beta}{8}}(r-r_{0})}{\cos^{3/5}\sqrt{\frac{5\alpha\beta}{8}}(r-r_{0})}.
\end{aligned}
\end{equation}
For $\beta < 0$, Eqs.(\ref{eq:Einstein-eqs1})-(\ref{eq:Einstein-eqs3})
 again admit two different types of solutions;
the warped solutions
\begin{equation}
\begin{aligned}
	&M_{w}^{0}(r) = C_{1}\exp\left( \pm\sqrt{\frac{-\alpha\beta}{10}}r \right),\\
	&L_{w}^{0}(r) = C_{2}\exp\left( \pm\sqrt{\frac{-\alpha\beta}{10}}r \right),
\end{aligned}
\end{equation}
and the divergent solutions
\begin{equation}
\begin{aligned}
	&M_{d}^{0}(r) = C_{1}\sinh^{2/5}\sqrt{\frac{-5\alpha\beta}{4}}(r-r_{0}),\\
	&L_{d}^{0}(r) = C_{2}\cosh^{2/5}\sqrt{\frac{-5\alpha\beta}{4}}(r-r_{0}).
\end{aligned}
\end{equation}

\subsubsection{Inflating branes case: $\gamma \ne 0$}
From now on we discuss the inflating branes case, \emph{i.e.} $\gamma \ne 0$.
We start with a differential equation derived from Eqs.(\ref{eq:Einstein-eqs1})-(\ref{eq:Einstein-eqs3}), given by
\begin{equation}
	M'^{2} =
	\frac{\gamma^{2}M^{3}-\frac{\alpha\beta}{10}M^{5}+C}{M^{3}}
	\label{eq:start-ode}
\end{equation}
where $C$ is an integration constant.

Equation (\ref{eq:start-ode}) admits two possibilities; $M' = 0$ and $M' \ne 0$.

\paragraph{The case of $M' = 0$.}
This case is equivalent to $M = \text{const} \equiv M_{C}$.
In this case, equations \eqref{eq:start-ode} and \eqref{eq:useful-relation} are not valid anymore and we have to consider the full set of Einstein equations. 
Doing so, we find
\be
M(r) = \sqrt{\frac{6 H^2}{\Lambda}} ,\ L(r)= \sqrt{\frac{2}{\Lambda}}\sin \left(\sqrt{\frac{\Lambda}{2}}r\right),
\ee
which can be understood as follows: the four dimensional slice have a positive curvature, we add extradimensionw with positive curvature as well, the curvature of the total spacetime is still positive and related to the positive cosmological constant. Note indeed that this solution is well defined only for positive bulk cosmological constant.

%If we specially choose $C = 0$, we find
%$M_{C=0} = \pm\sqrt{10\gamma^{2}/\alpha\beta}$ for the only case of $\beta > 0$.
%Since the relation (\ref{eq:useful-relation}) doesn't work in the case of $M' = 0$,
%we must directly compute the Einstein equation Eqs.(\ref{eq:Einstein-eqs1})-(\ref{eq:Einstein-eqs3}) to obtain solutions of %$L(r)$.
%From the four-dimensional component of Eqs.(\ref{eq:Einstein-eqs1})-(\ref{eq:Einstein-eqs3}),
%we have a differential equation given by
%\begin{equation}
%	(\ln L)'' + (\ln L)'^{2} - \frac{3\gamma^{2}}{M_{C}^{2}} = -\alpha\beta.
%	\label{eq:start-ode2}
%\end{equation}
%It is remarkable that Eq.(\ref{eq:start-ode2}) has a warped solution.
%If $(\ln L)'' = 0$, we have such the solution
%\begin{equation}
%	L_{w}(r) = C_{\pm}\exp\left(\pm\sqrt{\frac{3\gamma^{2}}{M_{C}^{2}} -\alpha\beta}\,r\right)
%	\label{eq:warp}
%\end{equation}
%for $3\gamma^{2}/M_{C}^{2} -\alpha\beta \ge 0$.
%Let us note that the special case of $C = 0$ doesn't admit the warped solution (\ref{eq:warp})
%since $3\gamma^{2}/M_{C=0}^{2} -\alpha\beta = -7\alpha\beta/10 < 0$.
%For $(\ln L)'' \ne 0$ we have two solutions for $3\gamma/M_{C}^{2} -\alpha\beta \lesseqgtr 0$;
%$$\begin{aligned}
%	&L_{d}^{M'=0}(r) = C_{1}\sinh\sqrt{\frac{3\gamma}{M_{C}^{2}} -\alpha\beta}(r-r_{0}),\\
%	&L_{p}^{M'=0}(r) = C_{2}\cos\sqrt{\alpha\beta - \frac{3\gamma}{M_{C}^{2}}}(r-r_{0}).
%\end{aligned}$$

\paragraph{The case of $M' \ne 0$.}
From Eqs.(\ref{eq:Einstein-eqs1})-(\ref{eq:Einstein-eqs3}) we find a solution of $M(r)$ as a quadrature and a solution of $L(r)$ as follows \cite{Brihaye:2006pi}:
\begin{equation}
\begin{aligned}
	&r - r_{0} = \int dM \sqrt{\frac{M^{3}}{\gamma^{2}M^{3}-\frac{\alpha\beta}{10}M^{5}+C}}\,,\\
	&L(r) = C_{L}\frac{dM}{dr}
		= C_{L}\sqrt{\frac{\gamma^{2}M^{3}-\frac{\alpha\beta}{10}M^{5}+C}{M^{3}}}\,.
\end{aligned}
\end{equation}
Here we limit the study to the case $C = 0$, the case with $C\neq0$ involves ellitic functions.
For $\beta = 0$ we have a cigar-type set of solutions given by
\begin{equation}
	M_{c}(r) = \gamma(r - r_{0}),\;\;\;
	L_{c}(r) = L_{0} \equiv \gamma\,C_{L}.
\label{bginf0}
\end{equation}
For $\beta > 0$ we again have periodic solutions
\begin{equation}
\begin{aligned}
	&M_{p}(r) = \sqrt{\frac{10\gamma^{2}}{\alpha\beta}}
		\sin\left(\sqrt{\frac{\alpha\beta}{10}}(r - r_{0})\right),\\
	&L_{p}(r) = L_{0}\cos\left(\sqrt{\frac{\alpha\beta}{10}}(r - r_{0})\right).
\end{aligned}
\label{bginf1}
\end{equation}
For $\beta < 0$ we have diverging solutions
\begin{equation}
\begin{aligned}
	&M_{d}(r) = \sqrt{\frac{10\gamma^{2}}{-\alpha\beta}}\sinh\left(\sqrt{\frac{-\alpha\beta}{10}}(r - r_{0})\right),\\
	&L_{d}(r) = L_{0}\cosh\left(\sqrt{\frac{-\alpha\beta}{10}}(r - r_{0})\right).
\end{aligned}
\label{bginf-1}
\end{equation}

\section{Asymptotic solutions}
\subsection{Near origin development}

The near origin behaviour of the functions $f,L,M$ subject to the boundary conditions \eqref{eq:BC-f} and \eqref{eq:BC-gr} is given by
\bea
f(r)&=& -(m-1)\pi + f^{(n)}(0) r^n/n! + \Ord{r}{n+1} \label{asym0f}
\\
L(r)&=&r+\frac{-2\gamma-\alpha\left(\beta+\mu-{\left(-1\right)}^m\mu\right)}{4}\frac{r^3}{3!}+\Ord{r}{4}\label{asym0l}
\\
M(r)&=&1+\frac{2\gamma-\alpha\left(\beta+\mu-{\left(-1\right)}^m\mu\right)}{4}\frac{r^2}{2!}+\Ord{r}{3},\label{asym0m}
\eea
where $f^{(n)}$ stands for $n^{\mbox{th}}$ derivative of $f$.

Note that higher order corrections are straightforward to compute.

\subsection{Large $r$ limit}
In this section, we give the leading asymptotic correction to the functions $f,M,L$. We focus on the case $\gamma\neq0$. The asymptotic solution is then given by \eqref{bginf0}, \eqref{bginf-1} and \eqref{bginf1} according to the sign of $\beta$. Note however that we couldn't find the subleading corrections in the case $\beta>0$; in this case, the metric functions are given to the leading order in terms of trigonometric functions; it is not possible to neglect such terms.

\subsubsection{$\beta <0$}
Considering the topological vacuum solution plus a fluctuation and suppressing subdominant terms in the equations, we get
\bea
\delta f&\approx& f^-_1 e^{ -\frac{r}{4}\left(\sqrt{-10\alpha\beta} +\sqrt{-10\alpha\beta+16\mu}\right)} 
+f^-_2 e^{ -\frac{r}{4}\left(\sqrt{-10\alpha\beta} -\sqrt{-10\alpha\beta+16\mu}\right)}\label{asym<0f}
\\
\delta M&\approx& M^-_1 e^{\sqrt{-\alpha\beta/10}r} + M^-_2 e^{-\sqrt{-8\alpha\beta/5}r},\label{asym<0l}
\\
\delta L&\approx& L^-_1 e^{\sqrt{-\alpha\beta/10}r} + L^-_2 e^{-\sqrt{-8\alpha\beta/5}r},\label{asym<0m}
\eea
where $\delta M$ and $\delta f$ denote the fluctuation around $M(r)$ and $f(r)$ such that $f(r) = -(m-1)\pi + \delta f(r),\ M(r)=\sqrt{-10\gamma/(3\alpha\beta)}\sinh(\sqrt{-\alpha\beta/10}r) + \delta M(r)$; and where $f^-_1,f^-_2,M^-_1,M^-_2,L^-_1,L^-_2$ are arbitrary constant. 

Clearly, we are looking for the modes with $M^-_1= f^-_2 =L^-_1=0 $ such that $\delta f$ and $\delta M$ are indeed fluctuations.

\subsubsection{$\beta=0$}
Once again, we start from the topological vacuum plus a fluctuation. Here it is possible to solve for the flutcuations without further assumptions:
\bea
\delta f &=& f^0_1 x^{-\frac{3}{2}}K_{\frac{3}{2}}(x),\\
\delta M &=& M^0_1 + \frac{M^0_2}{r^2},\\
\delta L &=& L^0_1 + \frac{L^0_2}{r^2},
\label{asyminf}
\eea
where $x = r\sqrt{\frac{n^2}{L_0^2\gamma} + \mu}$, $K_{n}(x)$ is the modified Bessel function of second kind and $f^0_1, M^0_1, M^0_2, L^0_1,L^0_2$ are arbitrary constants. In this case, we are interested to solutions with $L^0_1=M^0_1=0$.

Note that for large $r$, the function $f$ decays as $\frac{e^{-r\sqrt{\frac{n^2}{L_0^2\gamma} + \mu}}}{r^2}$.

\section{Numerical solutions}
We solved the system of ordinary differential equations numerically with the solver Colsys \cite{colsys} 
for many values of the parameters. Due to the large number of parameters, we decided to adopt the following approach: we keep fixed the value of the gravitational coupling, the bulk cosmological constant and the winding numbers. 
Then we vary the Hubble factor for different values of the strength of the potential. Note that the model without inflating 4D slices has been studied \cite{Kodama:2008xm}; the way we treat the problem allows a direct visualization of the influence of the Hubble parameter on the pattern of solutions.

Before discussing the case $\gamma\neq0$, we shortly remind the solutions obtained in ref \cite{Kodama:2008xm}. Essentially two types of solutions were found for any sign of the bulk cosmological constant, for instance, in the case $\beta=0$ the branches of solutions were discovered, with asymptotic corresponding respectively to flat space (String branch) and one analogue to the Melvin universe (Melvin branch). In the following we will present families of solutions for $\gamma\neq0$, extrapolating between these two sets.

In the analysis, we focused on the surface energy of the skyrmion $E$ and its mean square radius $MSR$ defined as
\be
E = 2\pi \int_0^\infty T_0^0 L(r)dr\ \ ,\ \  MSR = \int_0^\infty r^2 f'(r)\sin^2f(r) dr,
\ee
where $T_0^0 = - \tau_0$, see \eqref{stresstensor}.
The mean square radius allows to caracterize the extension of the brane in the transverse direction; 
the more $MSR$ is small, the more the brane is localised. 

\begin{figure}[h]
	\centering
		\includegraphics[scale=.6]{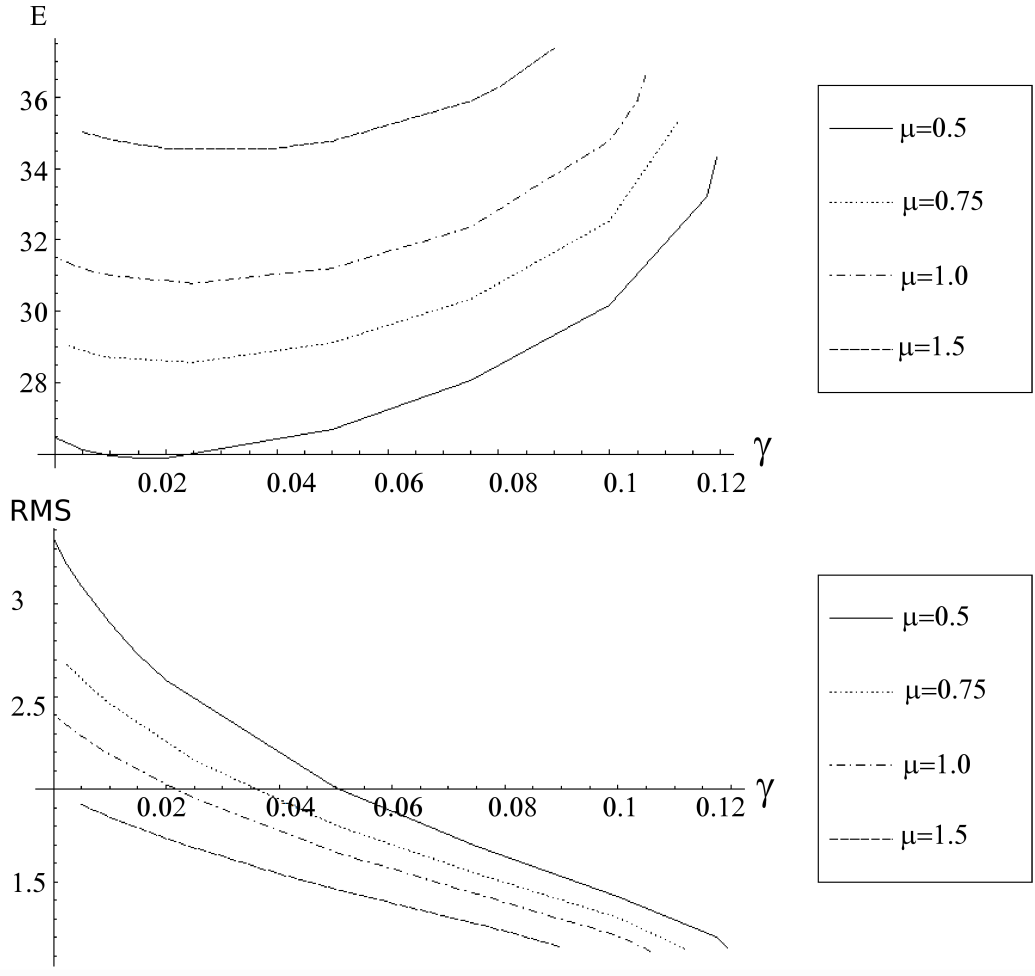}
	\caption{The value of the energy and of the mean square radius as a function of $\gamma$ for various values of $\mu$ and for $\alpha = 0.1,\ \beta=0.1$.}
	\label{fig:0101}
\end{figure}

\begin{figure}[h]
	\centering
		\includegraphics[scale=.65]{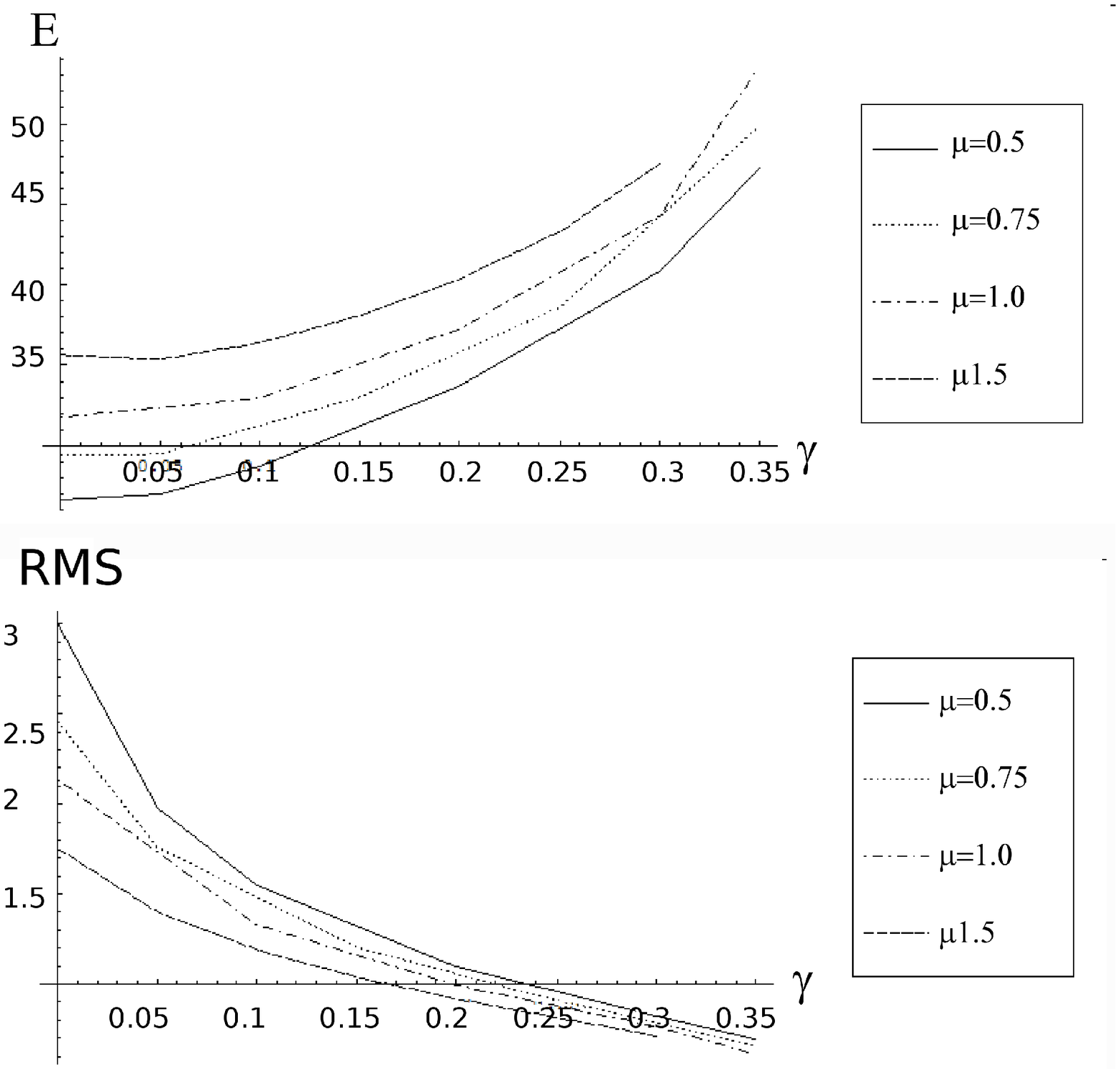}
	\caption{The value of the energy and of the mean square radius as a function of $\gamma$ for various values of $\mu$ and for $\alpha = 0.05,\ \beta=0.1$.}
	\label{fig:00501}
\end{figure}

Our results are summarized in figure \ref{fig:0101} and \ref{fig:00501} for 
$\alpha = 0.1,\ \beta=0.1$ and $\alpha = 0.05,\ \beta=0.1$ respectively and $m=n=1$. 
When the value of $\gamma$ increases for fixed $\alpha,\ \beta,\ \mu $ the mean square radius decreases 
while the energy decreases for small values of $\gamma$ and increases for larger values. In some intermediate values of $\gamma$, 
the energy of the skyrmion passes thought a minimum. The minimum occurs at smaller values of $\gamma$ when the value of $\mu$ is smaller. 
Note also that increasing values of $\mu$ leads to decreasing values of the mean square radius and decreasing values of the energy.

\begin{figure}[h]
	\centering
		\includegraphics[scale=.7]{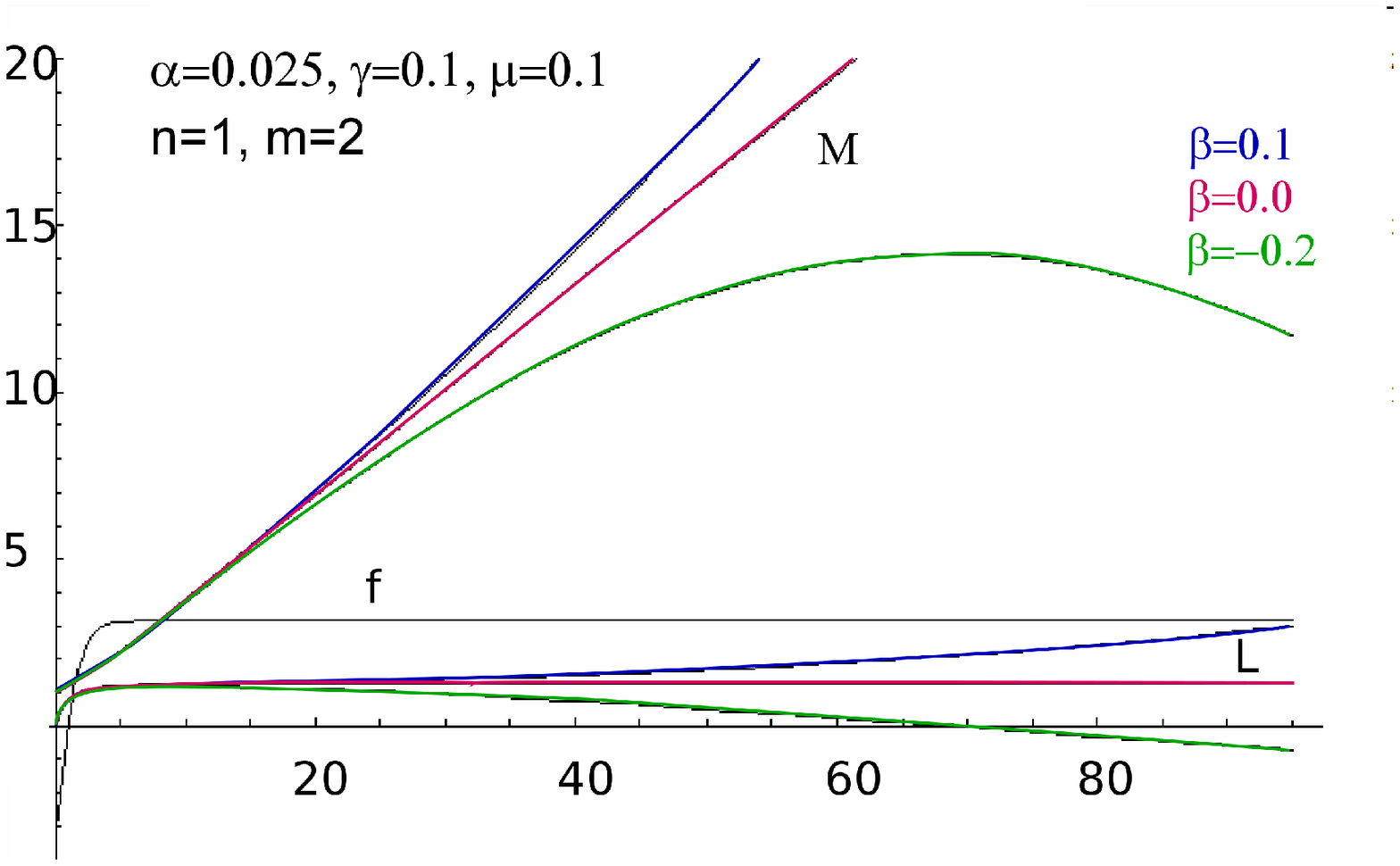}
	\caption{The typical profiles of the solutions for different signs of $\beta$ for $m=2$. The function $f$ tends quickly to its asymptotic value, so it is not possible to distinguish the different profiles for the function $f$.}
	\label{fig:prof m=2}
\end{figure}

\begin{figure}[h]
	\centering
		\includegraphics[scale=.7]{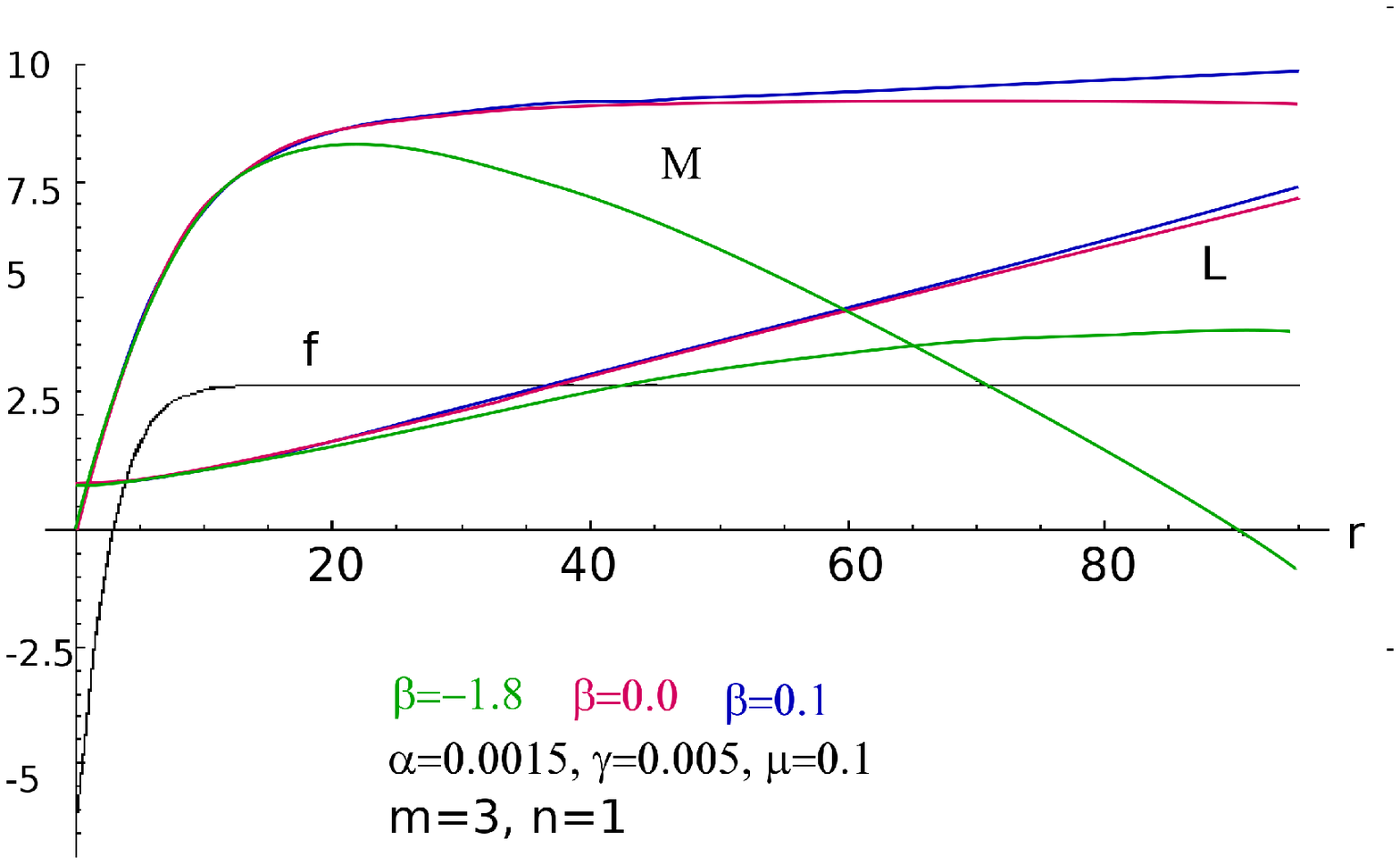}
	\caption{The typical profiles of the solutions for different signs of $\beta$ for $m=3$. Here again, it is not possible to distinguish the different profiles for the function $f$.}
	\label{fig:prof m=3}
\end{figure}

We present typical profiles of the solutions for $m=2,3$ in figure \ref{fig:prof m=2} and \ref{fig:prof m=3} 
respectively for non vanishing values of the parameters. These figures show that there are three possible geometries 
depending on the sign of the cosmological constant (see ref. \cite{Brihaye:2006pi}): opened ($\beta>0$), flat ($\beta=0$) and closed ($\beta<0$); all three geometries with angular deficits. This effect seems to be a generic feature of a model where the 4 dimensional branes are inflating. 

\begin{figure}[h]
	\centering
		\includegraphics[scale=.37]{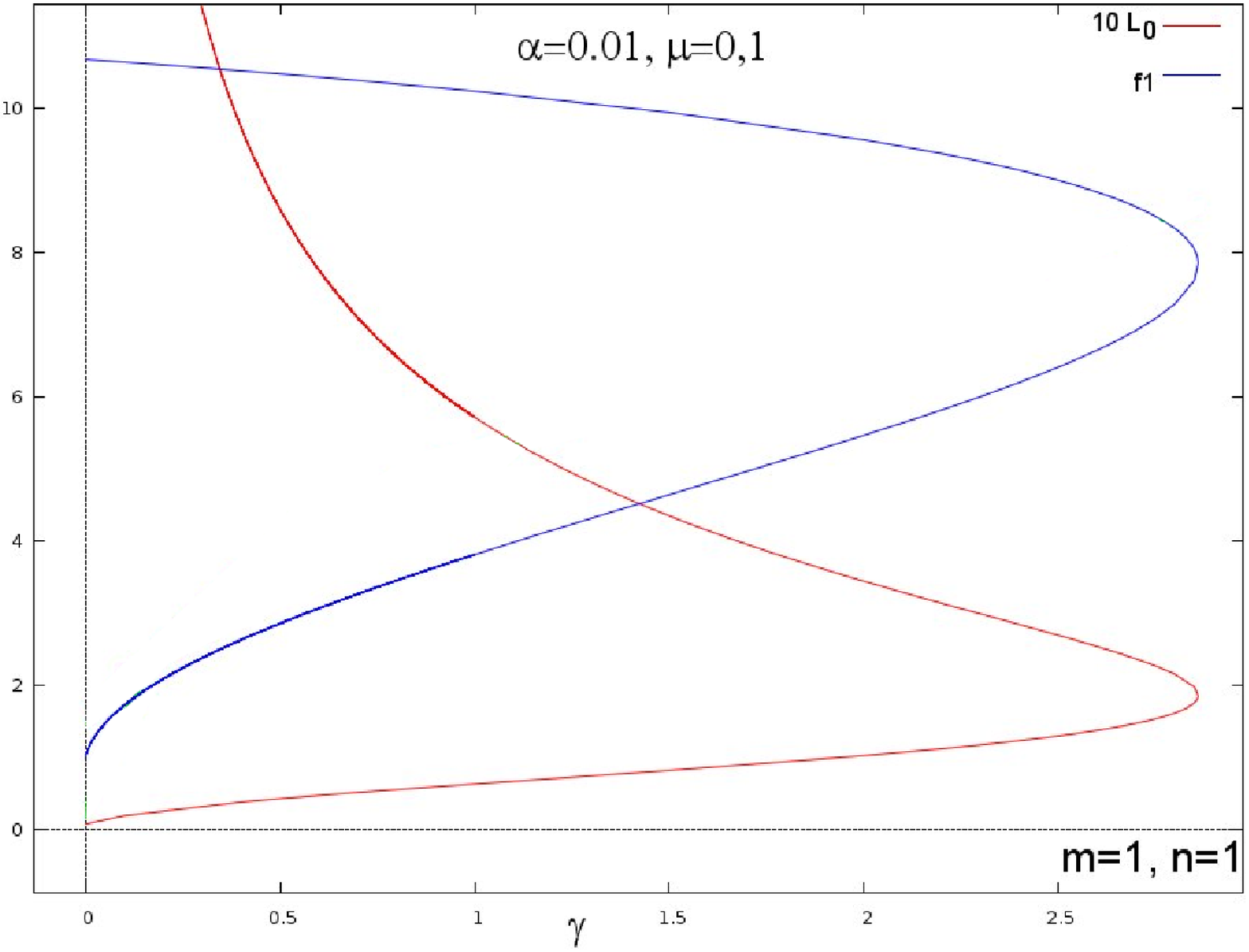}
	\caption{The value of the derivative at the origin of the baby Skyrme field, the coefficient $L_0$ of the metric function $L$ in \eqref{bginf0} for $\beta=0$. }
	\label{fig:2brb0}
\end{figure}

\begin{figure}[h]
	\centering
		\includegraphics[scale=.37]{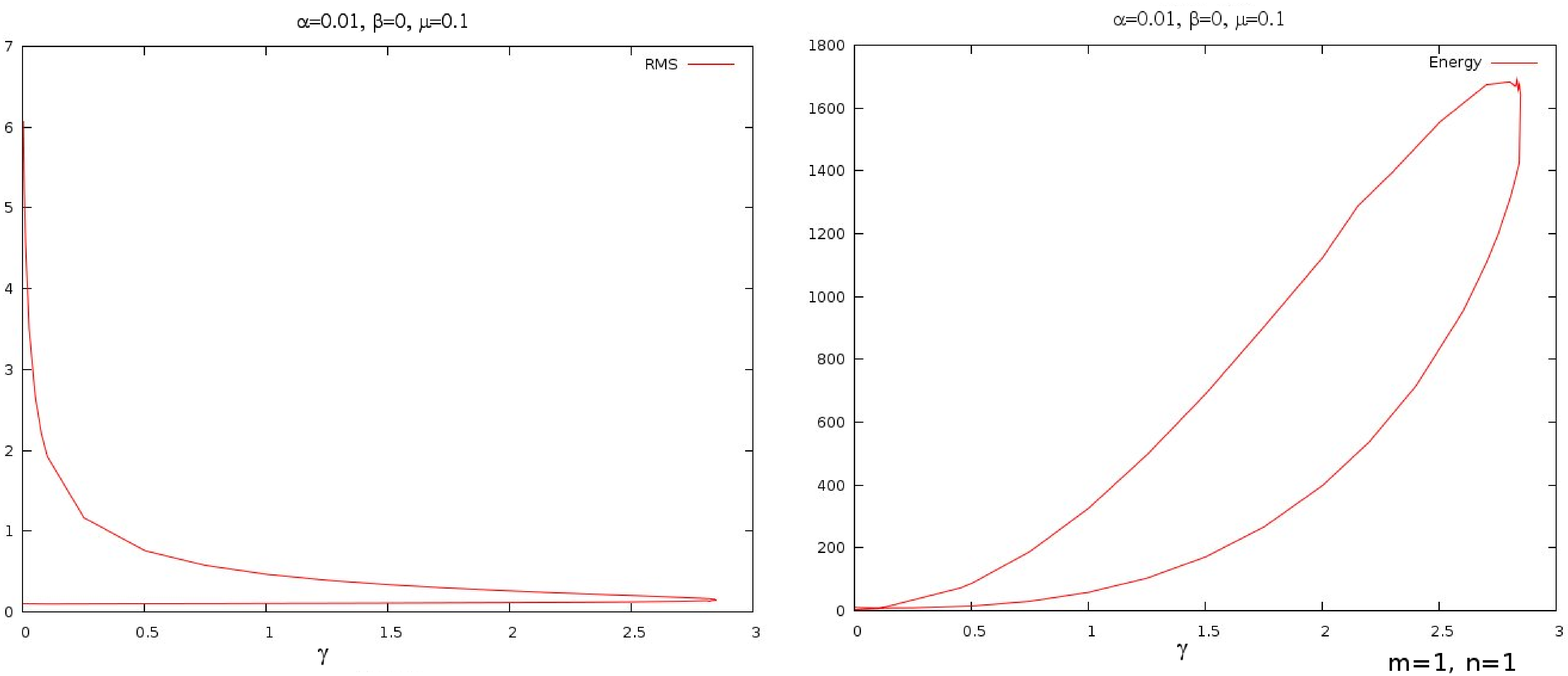}
	\caption{The value of the energy and of the square mean radius for the two branches with $\beta=0, m=1, n=1$.}
	\label{fig:ERmsb0}
\end{figure}

Pushing the investigation forward, it appears that the inflating baby-skyrmion exists up to a maximal value of the parameter $\gamma$ (the Hubble parameter). A second branch of solution exists as well for values of $\gamma$ lower than the maximal value. The second branch is of the same type as the first one and connects the two types of solutions available in the limit $\gamma\rightarrow0$. It should be mentioned that we start with the string solution in the case $\beta=0$ and increase the value of $\gamma$ until the second branch is reached, then we decrease $\gamma$ along the second branch. Figure \ref{fig:2brb0} and \ref{fig:ERmsb0} shows some relevant numerical parameters caracterizing the solution (resp. the energy and the square mean radius) with $\gamma\ne0$ and $\beta=0$.

\begin{figure}[h]
	\centering
		\includegraphics[scale=.3]{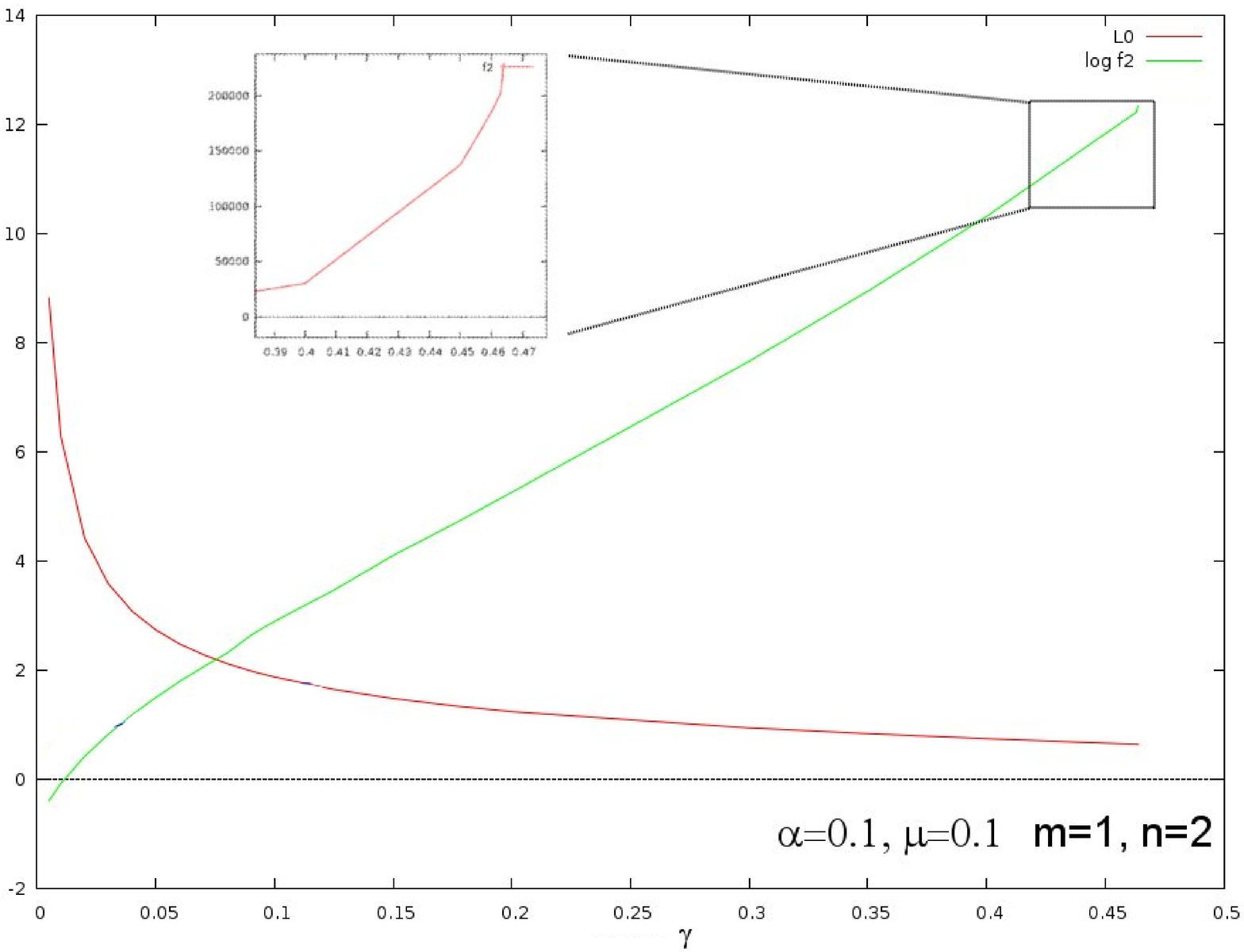}
	\caption{The asymptotic value of $L$ and the second derivative of $f$ at the origin (in log scale) for $\beta=0, m=1,n=2$. The second derivative $f_2$ diverges at the maximal value of $\gamma$. The window is a zoom if the region close to the maximal value of $\gamma$ showing $f_2$ (not in log scale).}
	\label{fig:n2crash}
\end{figure}

The case $n=2$ is however completely different. The solution still exists up to a maximal value of $\gamma$, but then the solution crashes; in the sense that the second derivative at the origin diverges at the maximal value of $\gamma$ (recall that the function $f$ behaves like $f\sim-(m-1)\pi + f_2 r^2$ close to the origin). This is shown in figure \ref{fig:n2crash} where the asymptotic value of $L$ and the second derivative of $f$ at the origin are shown. we did not find a second branch in this case; if such a branch exists, it seems unlikely that it will connect smoothly to the branch we constructed. 

\begin{figure}
\centering
	\includegraphics[scale=.5]{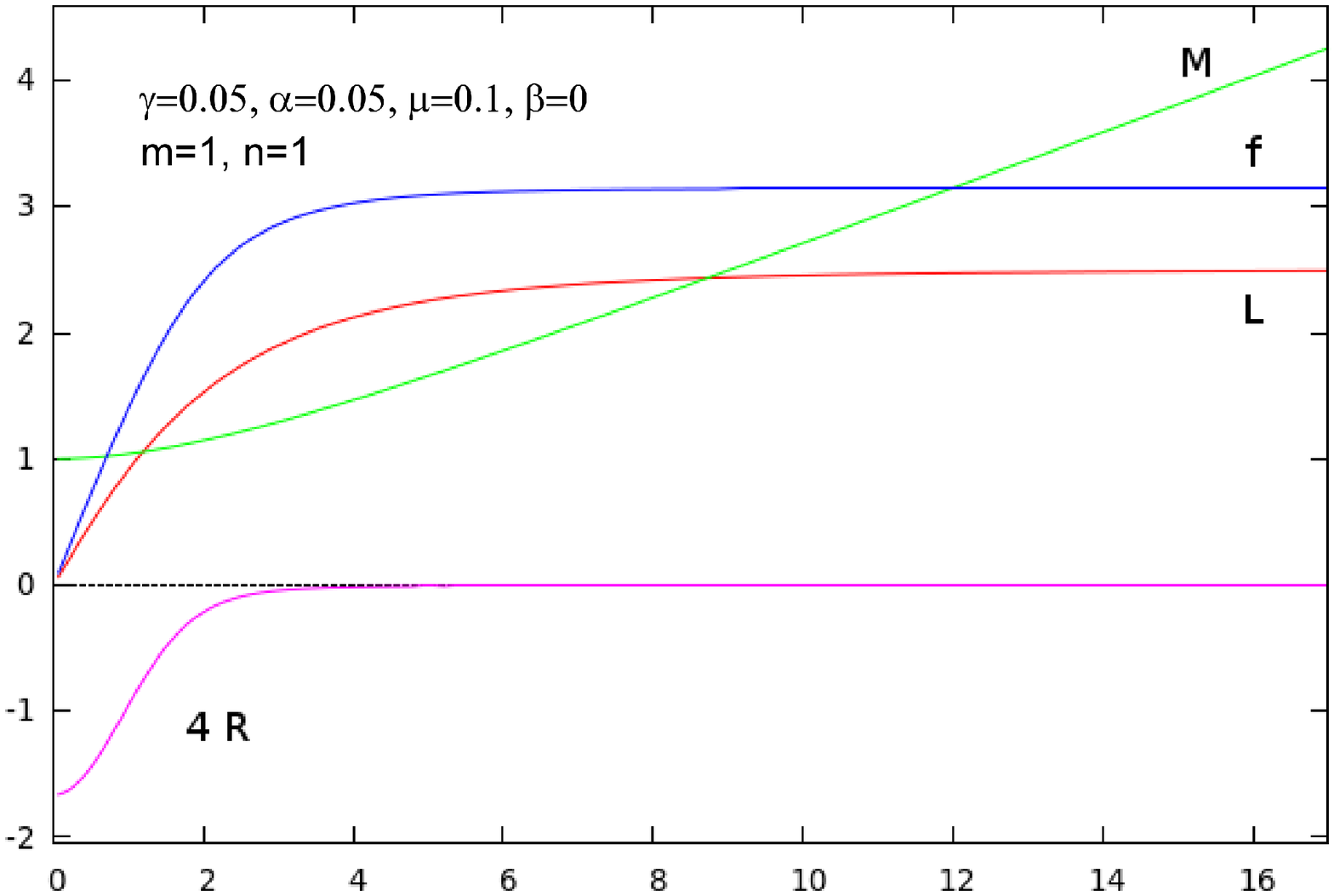}
	\caption{The typical profile of the metric functions, the baby-Skyrme field and the reduced scalar curvature for $n=1, m=1$.}
        \label{fig:Rn1}
\end{figure}

\begin{figure}
\centering
	\includegraphics[scale=.42]{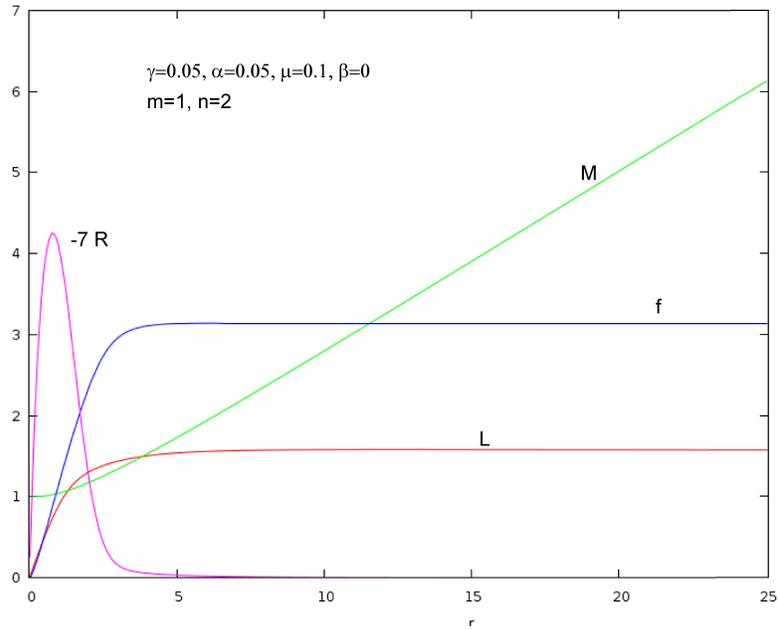}
	\caption{The typical profile of the metric functions, the baby-Skyrme field and the reduced scalar curvature for $n=2, m=1$.}
        \label{fig:Rn2}
\end{figure}

Note that the case $n=1$ and $n=2$ are also different from the geometrical point of view: the scalar curvature vanishes at the origin for $n=2$ while it goes to a non vanishing constant for $n=1$. This is illustrated on figure \ref{fig:Rn1} and \ref{fig:Rn2} for $\beta=0, n=1$ (resp $n=2$) where we show the metric functions, the baby-Skyrme function and the scalar curvature; the picture is similar for non vanishing values of $\beta$.

\clearpage

\section{Stability of the inflating baby-Skyrme brane}
In this section, we consider a general perturbation $h_{AB}$ around the background metric:
\begin{equation}
 ds^2 = M(r)^2\left( dt^2 - \exp(2Ht)d\vec x^2\right) - dr^2 - L(r)^2 d\theta^2 + h_{AB}(x^A) dx^A dx^B,
\end{equation}
and we choose the transverse traceless gauge $\nabla_Ah^{AB} = h_A^A = 0$. Note that $h^{AB} = g^{AC}g^{BD}h_{CD}$, $g$ being the background metric, and the inverse perturbed metric is given to order one in $h$ by $g^{AB}-h^{AB}$.

We will focus on the case $\beta=0$, already quite involved.

We parametrize the metric perturbation according to 
\begin{equation}
 h_{AB} = e^{\Omega t - i N \theta}\left(\begin{array}{ccccc}
                M(r)^2 H_{11}(r) & H_{12}(r)              &\ldots   & H_{15} (r) & H_{16}(r)\\
		H_{12}(r)        & M(r)^2e^{2Ht}H_{22}(r) &\ldots   & H_{25} (r) & \vdots\\
		\vdots           & \vdots                 &\ddots   & \vdots     & \vdots\\
		\vdots           & \vdots                 &         & H_{55} (r) & \vdots \\
		H_{16}(r)        & \ldots                 &\ldots   & \ldots     & L(r)^2H_{66}(r)
                \end{array}\right).
\end{equation}

The gauge conditions, $\nabla_Ah^{AB} = h_A^A = 0$ reduce to
\bea
&& H_{11}+H_{22}+H_{33}+H_{44}+H_{55}+H_{66}=0,\\
&& \frac{1}{L^2 M^4}\left(L M^2 L' H_{5j}+L^2 \left(M \left(4 M' H_{5j}+M H_{5j}'(r)\right)-(3 H+\Omega ) H_{1j}\right)-i N M(r)^2 H_{6j}\right)=0,\nonumber\\
&&\frac{1}{L^2 M^2} \left(L M^2 L'(H_{66}-H_{55})+L^2\left(M\left(M'H_{11}+M'H_{22}+M'H_{33}+M' H_{44}-4 M'H_{55}-MH_{55}'\right)\right.\right.\nonumber\\
&&\left.\left.-(3 H+\Omega ) H_{15}\right)-i N M^2 H_{56}\right)=0,\nonumber\\
&&\frac{1}{L^2 M^3} \left(-L M L' H_{15}+L^2 \left(M\left((3 H+\Omega ) H_{11}-H_{15}'-H H_{22}-H H_{33}-H H_{44}\right)-4 M' H_{15}\right)\right.\nonumber\\
&&\left.+i N M H_{61}\right)=0,\nonumber
\eea
where $j=2,3,4,6$. From the imaginary part of the gauge conditions, it follows that $H_{6A}=0,\ A=1,\ldots,6$. It is then possible to solve the gauge conditions for $H_{55}$ and $H_{5 m}'$, $m=1,\ldots,5$.

The matter field is parametrized as follows
\be
\vec n = (\sin( f(r) + \delta F(t,r,\theta))\sin n\theta, \sin( f(r) + \delta F(t,r,\theta))\cos n\theta,\cos( f(r) + \delta F(t,r,\theta)) ).
\ee
We parametrize the time and angular dependence according to
\be
\delta F(t,r,\theta) = e^{\Omega t - i N \theta}\phi(r). 
\label{pmatt}
\ee

In this setting the perturbed gravity equation in the transverse traceless gauge are given by
\be
-\frac{1}{2}(\Delta_L)_{ABCD}h^{CD}:=-\frac{1}{2}\left( \Box g_{AC}g_{BD} + 2 R_{ABCD}\right) h^{CD} = \alpha \left( \delta T_{AB} + \frac{1}{4}\left( \delta T g_{AB} + T h_{AB} - \Lambda h_{AB} \right) \right),
\label{pgrav}
\ee
where $T$ is the trace of the stress tensor, $\delta T_{AB}$ is the first order variation of the stress tensor due to the variation of the matter field and of the metric fieldsand where we restored dimensionless functions and parameters.

We fist work out the gravity equations. The component $(2,2), (3,3), (4,4)$ of the equations \eqref{pgrav} are formally the same for $H_{22}$ (resp. $H_{33},H_{44}$), allowing to set $H_{33}=H_{44}=H_{22}$. Furthermore, equations \eqref{pgrav} are compatible with 
\be
H_{12}=H_{13}=H_{14}=H_{23}=H_{24}=H_{25}=H_{34}=H_{35}=0,
\ee
by looking at the corresponding components of \eqref{pgrav}.

The surviving degrees of freedom are $H_{11}, H_{22}, H_{51}$. It turns out that it is possible to get a single equation out the components $(1,1), (2,2), (5,5), (5,1),(6,6)$ of \eqref{pgrav} by solving these equations for $H_{11}'',  H_{22}', H_{22}'', H_{51}$. In other words, we eliminate $H_{51}$ and it turns out that the equation for $H_{22}$ and for $H_{11}$ decouple. Note that solutions to \eqref{pgrav} are not pure gauge as long as the right hand side does not vanish.

We will consider the equation for $H_{11}$, given by
\be
-h''(r) + V_1(r) h'(r) + V_2(r) h(r) + S_1(r)\phi(r) + S_2(r) \phi'(r)=0,
\label{eqpertug}
\ee

\bea
V_1(r) &:=& -\frac{L'}{L} - 8 \frac{M'}{M},\\
V_2(r) &:=& \frac{n^2\alpha  f'^2 \sin ^2 f}{2 L^2}-\frac{1}{4} \alpha  f'^2-\frac{n^2 \alpha \sin^2f}{4 L^2}-\alpha  \mu  \cos f-\frac{2 L'
M'}{L M}+\frac{N^2}{L^2}-\frac{2 M''}{M}\\
&&-\frac{12 M'^2}{M^2}+\frac{3 \sqrt{\gamma } \omega }{M^2}+\frac{2 \gamma }{M^2}+\frac{\omega^2}{M^2}+\frac{\alpha  \beta }{2}-\alpha  \mu,\nonumber
\eea
where we use $h(r) = H_{11}(r)$ for shortness, $\omega = \sqrt{\frac{\kappa_4}{\kappa_2}}\Omega$ and where $S_1, S_2$ are involved combination of the background functions but are straightforward to compute.

The linearized equation for the matter fields is quite long and can be obtained from the variation of the reduced action, 
leading to an equation of the following form for the matter field perturbation:
\be
M^2\phi''(r) + P_1(r) \phi'(r) + (P_2(r) + m^2) \phi(r) + T_1(r) h(r)=0,
\label{eqpertum}
\ee
where 
\bea
&&P_1(r):=-\frac{M^2}{n^2\sin^2 f+L^2}\biggl[(N^2-3)n^2f'\sin 2f+\frac{L'}{L}(n^2\sin^2 f-L^2)
-4\frac{M'}{M}(n^2\sin^2 f+L^2)\biggr] \nonumber \\ \\
&&P_2(r):=-\frac{M^2}{n^2\sin^2 f+L^2}\biggl[n^2\frac{L'}{L}f'\sin 2f
+n^2\cos 2f\bigl((N^2-3)f'^2+N^2-1\bigr) \nonumber \\
&&\hspace{4cm}-n^2\sin 2f\Bigl(f''+4f'\frac{M'}{M}\Bigr)
-\mu L^2(N^2-1)\cos f\biggr]\nonumber\\
&&T_1(r):=\frac{H  (5 \sqrt{\gamma}+2 \omega ) \left(2 M(r)^2+1\right) f'(r)}{2 M(r)^3 M'(r)},
\label{equationp}
\eea
where we defined $m^2:=- \omega(3\sqrt{\gamma}+\omega)$ (see the following discussion).

Note that the system \eqref{eqpertug} and \eqref{eqpertum} constitutes an eigenvalue-like problem, the eigenvalue being essentially given by $\omega$.

In fact, $m^2$ appears as the (dimensionless) four-dimensional masses of the scalar four dimensional harmonics $\psi_m$:
\be
\nabla_\mu\nabla^\mu \psi_m=(\partial_t^2 + 3\sqrt{\gamma} \partial_t - \triangle) \psi_m = -m^2 \psi_m\,.
\label{4Dharmo}
\ee
Assuming the particular parametrisation of the perturbations (they don't depend on the four dimensional spatial coordinates), $m^2$ indeed reduces to $- \omega(3\sqrt{\gamma}+\omega)$.

In this case, we easily find the solution of \eqref{4Dharmo}:
\be
\psi_m \sim e^{\bigl(-\frac{3H}{2}\pm\sqrt{\left(\frac{3H}{2}\right)^2 - m^2}\bigr)t}.
\ee
It follows that the relevant parameter for the stability is $m^2$; modes with $m^2<0$ leads to tachyonic instabilities 
while modes with $m^2>0$ are stable.

\subsection{\label{4a}Numerical analysis: The formal discussion}
The most obvious boundary conditions are the conditions at infinity, where all the perturbations should vanish:
\be
\phi(\infty) = h(\infty) = 0.
\ee
In order not to spoil the topological properties of the baby skyrmion, we need to fix $\phi(0)=0$.
Close to the origin, the function $h$ should behave like $h\approx r^N$ (the detail of the asymptotic equation is the same as in the next section).

In practice, we integrate equations \eqref{eqpertug}, \eqref{eqpertum} in the following way:
first, we integrate the equations between $0$ and some intermediate values, say $r_m$, 
using a Runge-Kutta algorithm at order $4$, with the following boundary conditions: 
$\phi(0)=0,\ \phi'(0)=\delta_1,\ h'(0)=0,\ h(0)=h_0$ for some real values of 
$\delta_0, h_0$, then we integrate the equations backwards from a 
large value of the radial coordinates, say $r_l$ to $r_m$, 
imposing the suitable decay of the functions $\phi,\ h$ and their derivatives, given later. 
We generate two set of two linearly independent solutions, 
one between $0$ and $r_m$, say $h^1_L,\phi_L^1$ and $h^2_L,\phi_L^2$, the second between $r_m$ and $r_l$, 
say  $h^1_R,\phi_R^1$ and  $h^1_R,\phi_R^1$. The general solution is given by
\bea
h(r) &=& \left\{ \begin{array}{l}
                  a h^1_R + b h_R^2, \mbox{ for }r<r_m,\\
		  c h^1_L + d h^2_L, \mbox{ for }r>r_m,
                 \end{array}
		\right.\\
\phi(r) &=& \left\{ \begin{array}{l}
                  a \phi^1_R + b \phi_R^2, \mbox{ for }r<r_m,\\
		  c \phi^1_L + d \phi^2_L, \mbox{ for }r>r_m,
                 \end{array}
		\right.
\eea
where $a,b,c,d\in\mathbb R$. The solution is smooth if the values of the function and of their derivative match at the intermediate value $r_m$. The matching condition is expressed by
\be
\det \left(   
	\begin{array}{cccc}
	h^1_L(r_m)                & h^2_L(r_m)          & h^1_R(r_m)         & h^2_R(r_m)\\
	\phi^1_L(r_m)         & \phi^2_L(r_m)   & \phi^1_R(r_m)  & \phi^2_L(r_m)\\
	h^1_L\ '(r_m)               & h^2_L\ '(r_m)         & h^1_R\ '(r_m)        &  h^2_R\ '(r_m)  \\
	\phi^1_L\ '(r_m)        & \phi^2_L\ '(r_m)  & \phi^1_R\ '(r_m)  & \phi^2_R\ '(r_m)
	\end{array}\right) = 0.
\label{eqdet}
\ee

\label{decay}
We computed the decay of the functions $h,\ \phi$ by solving to the leading order the perturbation equations in the asymptotic region. In the case $\beta=0$, we find
\bea
\phi\sim F_1 \frac{1}{r^2}e^{-\bigl[\sqrt{(N^2-1)\bigl(\frac{1}{L_0^2} + \mu}\bigr)\bigr]r},~~
h\sim H_1\frac{1}{r^4}e^{-\frac{N}{L_0}r},
\eea
where $F_1,H_1$ are normalisation factor. (The detailed analysis will be shown in next subsection.)

In principle, we are able to integrate equations \eqref{eqpertug}, \eqref{eqpertum} 
based on the method described above, and actually have found some solutions. 
However, the numerical investigation was plagued by number of difficulties: 
\begin{itemize}
 \item The coefficients of the system of coupled differential equations are given in terms of numerically computed functions. These functions were however computed with a relative precision of order $10^{-5}$.
 \item Since we integrate the equations from $0$ to some maximal value of $r$, say $r_c$, the boundary condition imposed are not exact. We checked that the solution follows the correct decay for larger values of $r$ by integrating the solution from $r_c$ to $r_c+ \delta$, $\delta$ being a real number. It turned out that close to $r_c$, the decay was good, but the error due to the cutoff showed up further from $r_c$. However, varying $r_c$ did not influence much the eigenvalues.
\end{itemize}

Thus we admit that our results are acceptable as preliminary results which give us an initial guess of the true eigenvalues. 
We will present more details in a further publication and here, we will analyse in detail the stability of the matter sector (resp. gravitational sector) in a fixed gravitational (resp. matter field) background in terms of a slightly different scheme.

\subsection{Stability of the Baby-skyrmions with fixed gravitational background}
In this section, we shall solve the eigenequation (\ref{eqpertum}) where all the gravitational background fields remain unperturbed.
This means that terms involving $h$ are eliminated from (\ref{eqpertum}). The equation becomes decoupled from $h$. 
Since the equation involves coefficients to be evaluated numerically, we should of course solve it using numerical methods.
Before moving to the numerical resolution, we might be able to get some intuitions about sign of the eigenvalues; which would give indications on the stability properties of the solutions.

We follow a scheme employed for the analysis of the six-dimensional Abelian vortex \cite{Peter:2003zg}.
Eq.(\ref{eqpertum}) can be rewritten in the form of a zero-eigenvalue mode Schr\"{o}dinger equation for the function
\begin{eqnarray}
\psi(r):=\exp\Bigl[\frac{1}{2}\int^r\frac{P_1(s)}{M^2(s)}ds\Bigr]\phi(r)\,,
\end{eqnarray}
namely
\begin{eqnarray}
-\psi''+V(r)\psi=0\,,
\label{schroedingereq}
\end{eqnarray}
where the potential is
\begin{eqnarray}
V(r)=W(r)'+W^2(r)-\frac{P_2+m^2}{M^2}
,~~~~W(r)=\frac{P_1}{2M^2}
\end{eqnarray}
Eq.(\ref{schroedingereq}) can be understood in the context of so called the supersymmetric quantum mechanics 
\cite{Cooper:1994eh} where
the function $W(r)$ is identified as the superpotential in the SUSY QM. 
 The equation can be rewritten
 by using the operator 
\begin{eqnarray}
A=\frac{d}{dr}+W(r),~~A^\dagger=-\frac{d}{dr}+W(r)\,,
\end{eqnarray}
namely
\begin{eqnarray}
\Bigl(AA^\dagger-\frac{P_2+m^2}{M^2}\Bigr)\psi=0.
\end{eqnarray}
Formally, the lowest eigenvalue of the operator $A A^\dagger$ equals to the first excited state of the $A^\dagger A$, obtained by the reversing the order of the operator $A,A^\dagger$. If the ground state of the operator $A^\dagger A$ is a zero mode, i.e.$A^\dagger A\psi=0$, one easily see that the eigenvalues of the operator $AA^\dagger$ are positive-definite. Note however that the potential in the $A^\dagger A$ contains deep negative well, so has many negative eigenvalues. Note also that strictly speaking, $V_M$ is not exactly a suspersymmetric potential, due to the presence of the additial $(P_2+m^2)/M^2$ term.

\begin{figure}[H]
	\centering
		\includegraphics[scale=1.0]{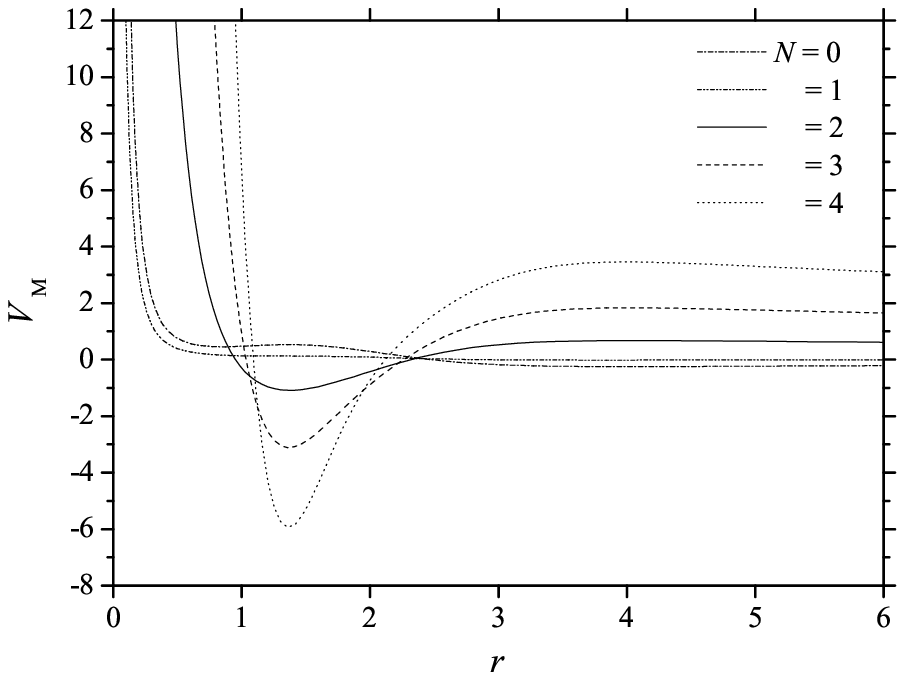}
	\caption{The potential $V_{\rm M}$ (\ref{effpotm}) for $n=1$ 
	with the parameter $\alpha=0.1, \beta=-0.15, \gamma=0.02, \mu=0.1$.}
	\label{fig:mass_potential}
\end{figure}

Here we plot the effective potential 
\begin{eqnarray}
V_{\rm M}(r):=W'(r)+W(r)^2-\frac{P_2(r)}{M^2(r)}\,.
\label{effpotm}
\end{eqnarray}
In order that (\ref{schroedingereq}) exhibits a zero-eigenvalue solution, the potential $V_{\rm M}-m^2/M^2$ should contains some negatives regions. In Fig.\ref{fig:mass_potential}, we plot the function $V_{\rm M}(r)$ for the first few of $N$ with a typical gravitational background. 
The case of $m^2>0$ is trivial; by suitably adjusting values of $m^2$ we could get the solutions. Those are the eigenvalues which might be observed at the LHC or some other probes.  
On the other hand, for $m^2<0$ the situation is more complicated. If the potential is positive-definite, apparently we have no solution of (\ref{schroedingereq}), thus the possibility of tachyonic mode is removed. 
From this point of view, the solutions for $N=0,1$ seem to have no tachyonic mode. However, the potentials for $N\ge 2$ have a negative pit at the core of the brane, which seems to grow as $N$ increase. 
Thus the tachyonic mode may occurs especially for larger $N$. For the case of $N=2$, the potential is shallow but still has a negative well; the equation might have a tachyonic mode. 
In order to go further, however, we have to rely on the numerical study. 

The eigenproblem can be solved numerically by the standard {\it predictor-corrector method}~\cite{p-c-method}.
The detail of the procedure is described in Appendix A. Here we shall examine informations of limiting 
behavior of the fluctuation $\phi$, but first, we will study the near origin and asymptotic behavior of the fluctuation $\phi$.

\subsubsection{Asymptotic behaviors}
At the vicinity of the origin, the asymptotic form of solution depends on the winding number $n$.
By using the asymptotic solution at the vicinity of the origin (\ref{asym0f})-(\ref{asym0m}), 
one can find the linearized equation of (\ref{eqpertum}) as
\be
\phi''+\frac{p_1}{r}\phi'+\frac{p_2}{r^2}\phi=0
\ee
which has a solution as the form:
\be
\phi(r)=c_+r^{\lambda_+}+c_-r^{\lambda_-},~~~~
\lambda_\pm=\frac{1}{2}\Bigl[(1-p_1)\pm\sqrt{(1-p_1)^2-4p_2}\Bigr]
\ee
where $c_\pm$ are arbitrary integration constants.
For $n=1$, the coefficients of the equation are 
\be
p_1=-\frac{1}{1+u^2}(2N^2u^2-5u^2-1),~~
p_2=N^2-1\,
\ee
where $u:=f'(0)$. After a slight examination, one finds the regular solution only the case for $N\ge 2$.
Similarly, for $n=2$ 
\begin{eqnarray}
p_1=1,~~p_2=-4(N^2-1)
\end{eqnarray}
so one easily see that for $N\ge 2$ the solutions are regular. 
As a conclusion, the solutions at the origin are regular for $N\ge 2$ but not for $N<2$ .

For large values of the radial coordinate, the behavior of the function $\phi$ essentially depends on the sign of $\beta:=\Lambda_6\kappa_4/\kappa_2^2$.
For $\beta>0$, the metric solutions are periodic. It is an unpleasant feature; the periodic behavior seems to interrupt us to find the good asymptotics.
Of course it does not necessarily indicate that the solution is unstable but, for the time being, we concentrate our analysis on the case $\beta\le 0$.

Recall that for $\beta=0$, (\ref{bginf0}),(\ref{asyminf}) leads to
\begin{eqnarray}
f\sim \pi,~~
M\sim \gamma r,~~L\sim L_0
\end{eqnarray} 
where $ L_0$ is a constant.
So, sufficiently far from the origin, equation (\ref{eqpertum}) becomes
\begin{eqnarray}
r^2\phi''+4r \phi'+\Bigl[-(N^2-1)\Bigl(\mu+\frac{n^2}{L_0^2}\Bigr)
r^2+\frac{m^2}{\gamma^2}\Bigr] \phi=0;~
\end{eqnarray}
Since this equation is a kind of Bessel's differential equation, 
the solutions can be written in terms of Bessel function with real/complex variables. 
The asymptotic solution is of the form:
\be
\phi(r)\sim \dfrac{1}{r^2}\exp
\Bigl(-\sqrt{(N^2-1)\Bigl(\frac{n^2}{L_0^2}+\mu\Bigr)}r\Bigr)
\ee
Thus we can find non asymptotically diverging modes of the function $\phi$ for all value of $N$. (Note that 
because of the condition at the origin, only $N\ge 2$ is acceptable. ) 

For $\beta<0$, from (\ref{bginf-1}), (\ref{asym<0f})-(\ref{asym<0m}), 
The asymptotic behavior of the solutions is 
\begin{eqnarray}
f\sim\pi,~~M\sim\dfrac{\gamma}{\xi}\sinh \xi r,~~L\sim L_0\cosh \xi r,~~~~
\xi:=\sqrt{\dfrac{-\alpha\beta}{10}}
\end{eqnarray}
The equation (\ref{eqpertum}) at the far from the origin is
\begin{eqnarray}
\sinh^2\xi r \phi''
+\xi(\sinh \xi r\cosh \xi r+4\cosh^2\xi r)\phi'
+\Bigl[-(N^2-1)\Bigl(\dfrac{n^2}{L_0^2}+\mu\cosh^2\xi r\Bigr)+
\dfrac{\xi^2m^2}{\gamma^2}\Bigr]\phi=0
\end{eqnarray}
For sufficient large $r$, the equation becomes
\begin{eqnarray}
\phi''+5\xi \phi'-\mu (N^2-1)\phi=0,
\end{eqnarray}
the localizing modes are thus 
\be
\phi (r)\sim \exp\biggl[\Bigl(-\frac{5}{2}\xi\pm\sqrt{\Bigl(\frac{5}{2}\xi \Bigr)^2+\mu (N^2-1)}~\Bigr)~r\biggr].
\ee
For $N\ge 2$ only the minus sign is available while $N=0$ both signs are acceptable.

\subsubsection{The numerical results}
\begin{figure}[H]
	\centering
		\includegraphics[scale=.8]{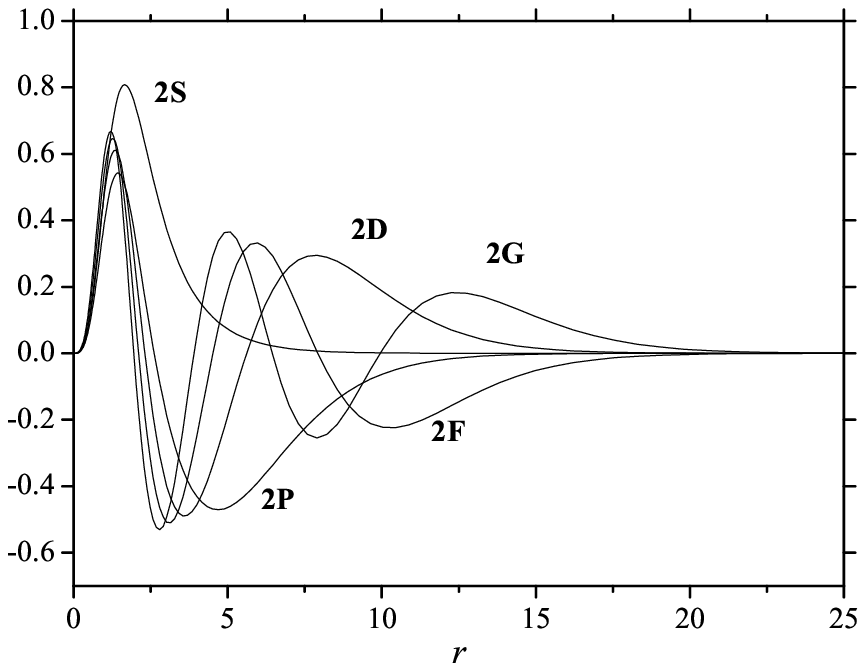}
		\includegraphics[scale=.8]{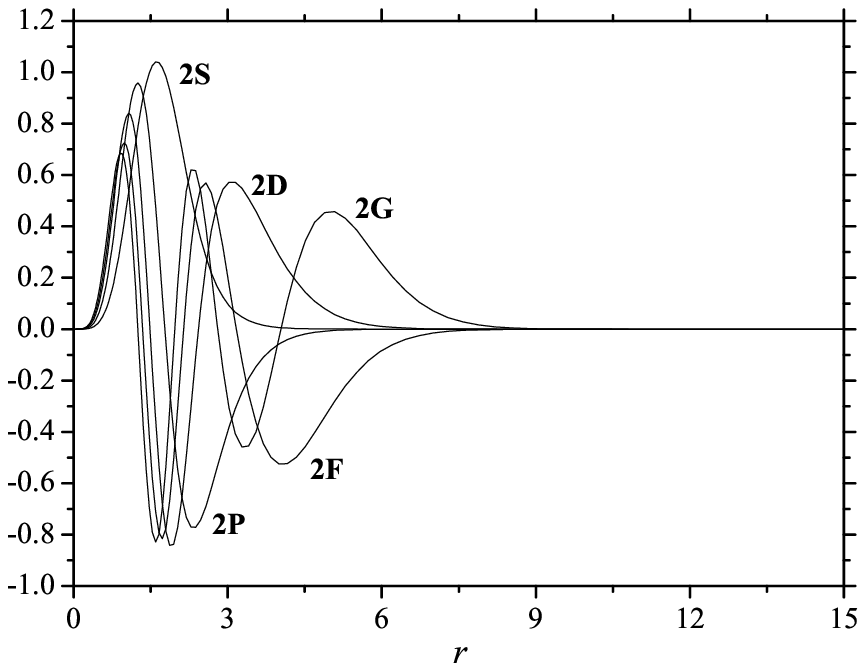}
	\caption{A first few of the fluctuations $\phi$. 
We denote $2{\rm S},2{\rm P},\cdots$ which indicate $N=2$ and the number of nodes. 
The background profiles are of the parameter $\alpha=0.1, \gamma=0.02, \mu=0.1$.
The left is for $(m,n)=(1,1)$, $\beta=0$. The right is for $(m,n)=(1,2)$, $\beta=-0.15$.}
	\label{fig:stabef}
\end{figure}

\begin{figure}[H]
	\centering
		\includegraphics[scale=1.0]{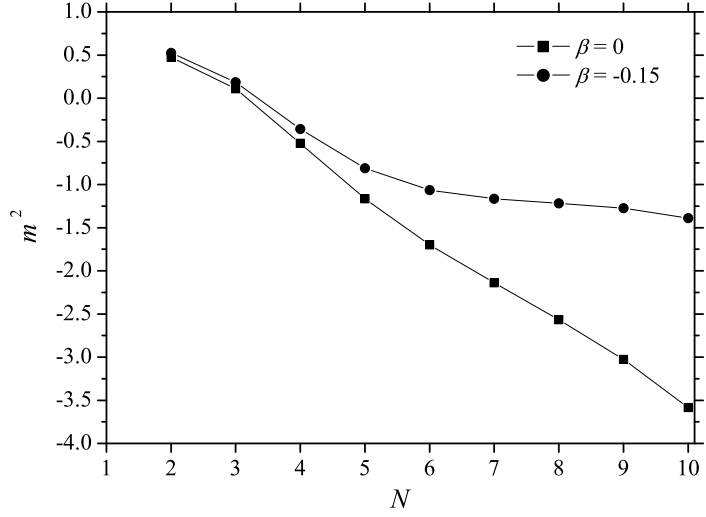}
	\caption{The lowest eigenvalues $m^2$ (with no-node) for $N=2\sim 10$ 
	with the parameter $\alpha=0.1, \beta=0, -0.15, \gamma=0.02, \mu=0.1$.}
	\label{fig:tachyonicmodes}
\end{figure}

For $N=2$, we have found a  tower of positive, discrete eigenvalues for both zero and 
negative $\beta$. 
In Fig.\ref{fig:stabef}, we plot some eigenfunctions $\phi$ of Eq.(\ref{eqpertum}) for 
$n=1,2$ and $\beta=0$ or $-0.15$. The corresponding eigenvalues are assembled in Table \ref{tab:stabev}.
For all cases we have many localized modes; there are essentially no notable differences for the changes of 
$\beta$ or $n$. On the other hand, for $N\ge 2$, the tachyonic modes may exist because of the deep negative potential.
In Fig.\ref{fig:tachyonicmodes}, we show the eigenvalues corresponding to the eigenfunction with no-node
for the parameter set:$\alpha=0.1, \beta=0, -0.15, \gamma=0.02, \mu=0.1$. 
At least, within this parameter set, we find that we  
unavoidably get the tachyonic modes except for $N=2,3$.

We also study the stability property of the second (``unstable'') branch. 
The eigenvalues are assembled in Table \ref{tab:stabev12}. For the second branch solutions, we got the highly 
localized eigenfunction for the fluctuation $\phi$; as a result, the eigenvalues are quite 
higher than the first branches.
Note that our obtained eigenvalues are dimensionless one. In order to recover the dimensionful one, 
we need information about the Skyrme parameters $\kappa_2,\kappa_4$.
In Ref.\cite{Kodama:2008xm}, 
we have estimated masses of fundamental fermions localized on the warped baby-skyrmion branes with 
negative cosmological constant. In rough speculation, we can extract the value of parameter as
$\sqrt{\kappa_2/\kappa_4}\sim 10^4$ MeV. From this we find
\begin{eqnarray}
\sqrt{-\Omega (3H+\Omega)}=\sqrt{\frac{\kappa_2}{\kappa_4}}m &\sim& 13~{\rm TeV} : {\rm the~first~branch}\nonumber \\
&\sim& 87~{\rm TeV} : {\rm the~second~branch}
\end{eqnarray}
for the data of Table \ref{tab:stabev12}.
Apparently, the first branch is in the TeV scales, but 
the second branch solution could not be observed in experimental facility.  

\subsection{Stability of the metrics with fixed skyrmion background}
If the background skyrmions remain unperturbed, the equation (\ref{eqpertug}) can easily 
be solved in terms of the method which has been described in the case of the baby-skyrmions.
This corresponds to omit the terms $\phi,\phi'$.  
In this case we define the eigenvalue $l^2:=-(\omega+\sqrt{\gamma})(\omega+2\sqrt{\gamma})$.

\subsubsection{Asymptotic behaviors}
Similar to the matter field case, at first we examine the asymptotic behaviors of the solutions. 
By using the asymptotic solution at the vicinity of the origin (\ref{asym0f})-(\ref{asym0m}), 
one can find the linearized equation of (\ref{eqpertug}) as
\be
h''+\frac{1}{r}h'-\frac{N^2}{r^2}h=0
\ee
which has a solution as the form $h(r)\sim r^{N}$. Thus, for $N\ge 1$ we have regular solutions while 
$N=0$ are not adequate for the regularity condition at the origin. 

Sufficiently far from the origin, one straightforwardly finds the form of the linearized 
equation of (\ref{eqpertug}) for $\beta=0$
\begin{eqnarray}
r^2h''+8r h'+\Bigl(12+\frac{l^2}{M_0^2}\Bigr)h-r^2\frac{N^2}{L_0^2}h=0;~
\end{eqnarray}
The asymptotic solution is of the form:
\be
h(r)\sim \dfrac{1}{r^4}\exp
\Bigl(-\frac{N}{L_0}~r\Bigr)
\ee
Thus we can find non asymptotically diverging modes of the function $h$ for all value of $N$. (Note that 
because of the condition at the origin, only $N\ge 1$ is acceptable. ) 

For $\beta<0$, we find the modes 
\begin{eqnarray}
h(r)\sim \exp
\Bigl[\Bigl(-\frac{9}{2}\omega\pm\sqrt{\frac{17}{4}\omega^2+\frac{\alpha\beta}{2}}~\Bigr)~r\Bigr]
\end{eqnarray}
This solution descreases as $r\to \infty$.

\subsubsection{The numerical results}
Contrary to the case of the baby-skyrmions, only for $N=2$, we have found a tower of positive, discrete eigenvalues for both zero and negative $\beta$. For $N=1$, the solution seems to be tachyonic. 

\begin{figure}[H]
	\centering
		\includegraphics[scale=.8]{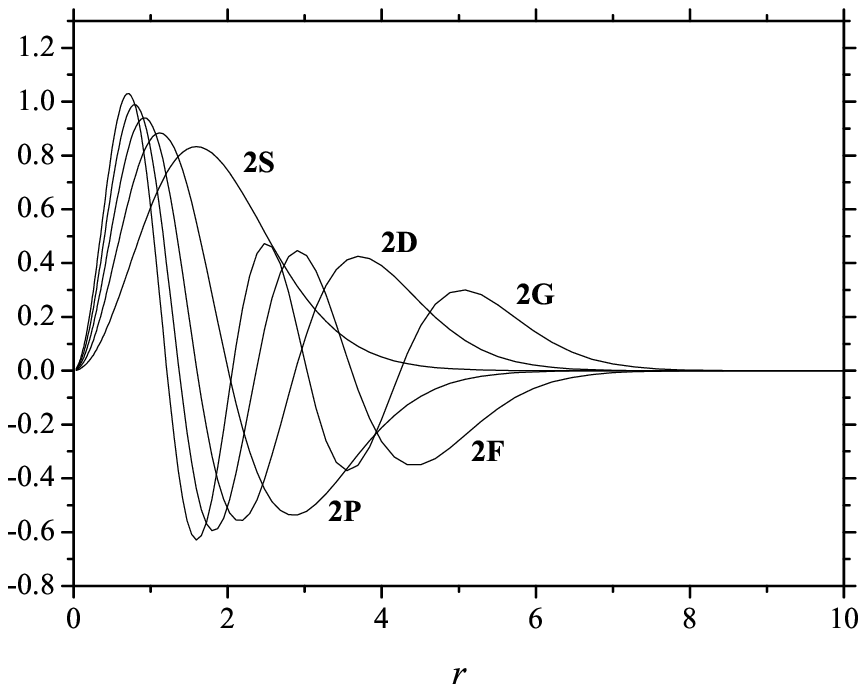}
		\includegraphics[scale=.8]{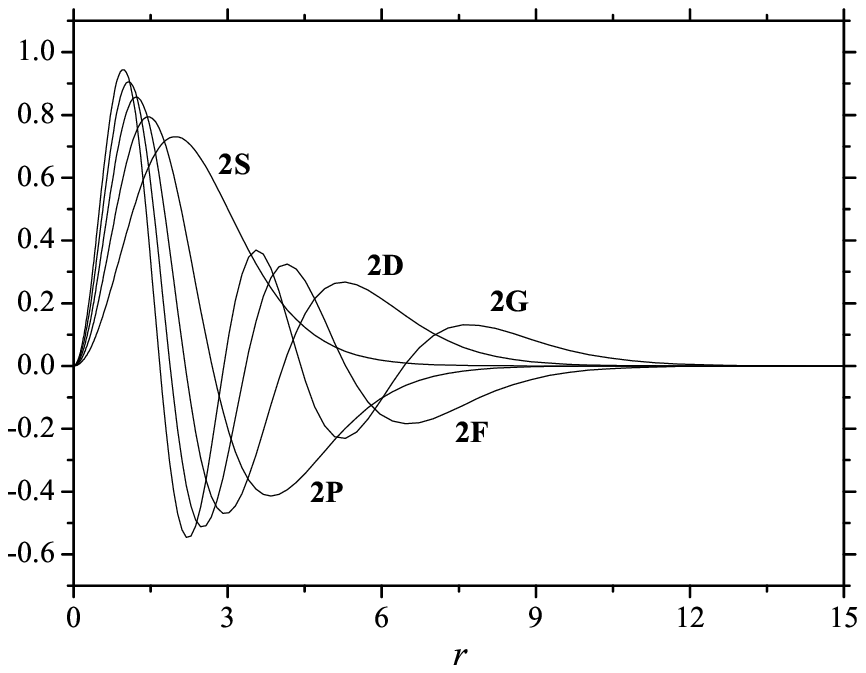}
	\caption{A first few of the fluctuations $h$. 
We denote $2{\rm S},2{\rm P},\cdots$ which indicate $N=2$ and the number of nodes. 
The background profiles are of the parameter $\alpha=0.1, \gamma=0.02, \mu=0.1$.
The left is for $(m,n)=(1,1)$, $\beta=0$. The right is for $(m,n)=(1,2)$, $\beta=-0.15$.}
	\label{fig:stabef_gravity}
\end{figure}

\begin{figure}[H]
	\centering
		\includegraphics[scale=1.0]{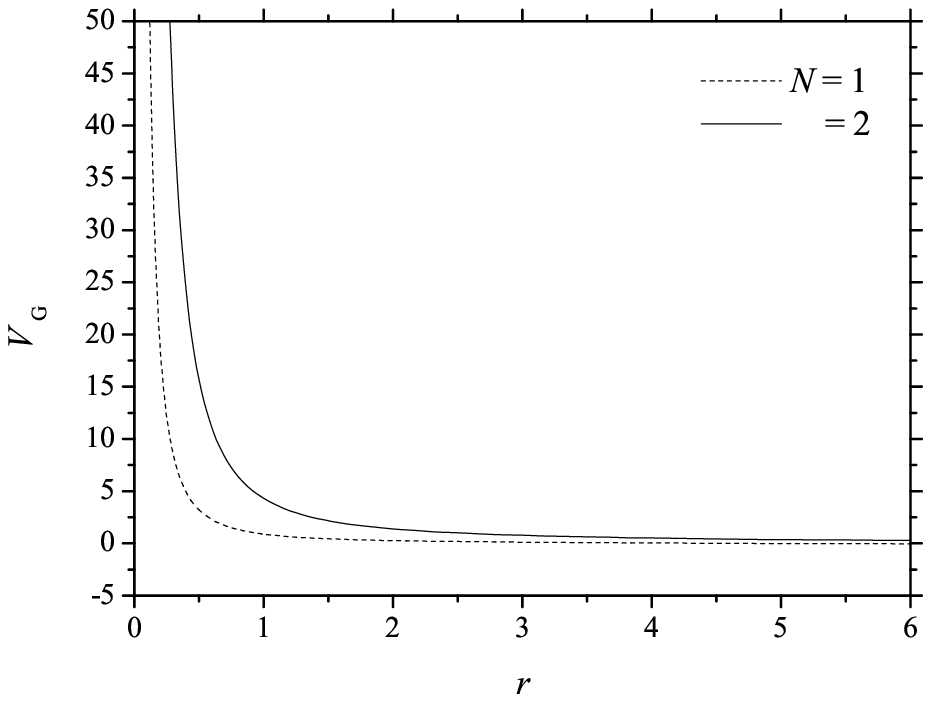}
	\caption{The potential $V_{\rm G}$ estimated by (\ref{efpotg}) for $n=1$ 
	with the parameter $\alpha=0.1, \beta=-0.15, \gamma=0.02, \mu=0.1$ and for $N=1,2$.}
	\label{fig:mass_potential_gravity}
\end{figure}

In Fig.\ref{fig:stabef_gravity}, we plot some eigenfunctions $\phi$ of Eq.(\ref{eqpertug}) for 
$n=1,2$ and $\beta=0$ or $-0.15$. The corresponding eigenvalues are assembled in Table \ref{tab:stabev_gravity}.
Similar to the case of the skyrmions, 
we plot the effective potential for the gravity in Fig.\ref{fig:mass_potential_gravity}
\begin{eqnarray}
V_{\rm G}(r):=W'(r)+W(r)^2-\frac{V_2(r)}{M^2(r)}\,.
~~W(r):=\frac{V_1(r)}{2M^2}
\label{efpotg}
\end{eqnarray}
For $N=1$, the potential has a subtle negatives, so it produces the tachyonic 
mode. On the other hand, the potential of $N=2$ is positive-definite, so the possibility of tachyonic mode is excluded
for the gravity perturbation in this case. 

\begin{table}[H]
\begin{tabular}{|c|c|c|c|c|} \hline
& \multicolumn{2}{|c|}{$n=1$} &
\multicolumn{2}{|c|}{$n=2$} \\\hline 
& $\beta=0$ & $\beta<0$ & $\beta=0$ & $\beta<0$ \\ \hline
~~~~S~~~~&~~0.4759~~&~~0.5264~~&~~1.4491~~&~~0.8249~~\\ \hline
~~~~P~~~~&~~1.5109~~&~~1.6485~~&~~8.8910~~&~~6.5876~~\\ \hline 
~~~~D~~~~&~~2.1979~~&~~2.4023~~&~~16.657~~&~~11.885~~\\ \hline
~~~~F~~~~&~~2.9344~~&~~3.2142~~&~~24.377~~&~~15.899~~\\ \hline 
~~~~G~~~~&~~3.7193~~&~~4.0860~~&~~31.362~~&~~19.390~~\\\hline
	     \end{tabular}~~~~~~
\caption{For the case of $N=2$, the eigenvalues $m^2$ of the equation (\ref{eqpertum}) for $(m,n)=(1,1),(1,2)$ with the parameter $\alpha=0.1, \gamma=0.02, \mu=0.1$, $\beta=0.0,-0.15$.}
\label{tab:stabev}
\end{table}

\begin{table}[H]
\begin{tabular}{|c|c|c|c|} \hline
 \multicolumn{2}{|c|}{$\beta=0$} &
\multicolumn{2}{|c|}{$\beta<0$} \\\hline 
$1_{\rm st}$ & $2_{\rm nd}$ & $1_{\rm st}$ & $2_{\rm nd}$ \\ \hline
~~1.7437~~&~~76.295~~&~~1.7475~~&~~76.293~~\\ \hline
\end{tabular}~~~~~~
\caption{For the case of $N=2$, the eigenvalues $m^2$ of the first and the second branches for $(m,n)=(1,1)$with the parameter $\alpha=0.01, \gamma=0.1, \mu=0.1$, $\beta=0.0,-0.1$.}
\label{tab:stabev12}
\end{table}

\begin{table}[H]
\begin{tabular}{|c|c|c|c|c|} \hline
& \multicolumn{2}{|c|}{$n=1$} &
\multicolumn{2}{|c|}{$n=2$} \\\hline 
& $\beta=0$ & $\beta<0$ & $\beta=0$ & $\beta<0$ \\ \hline
~~~~S~~~~&~~0.7591~~&~~0.7664~~&~~5.2498~~&~~3.4840~~\\ \hline
~~~~P~~~~&~~1.2026~~&~~1.0747~~&~~8.0611~~&~~5.2452~~\\ \hline 
~~~~D~~~~&~~1.7024~~&~~1.3397~~&~~12.049~~&~~6.9922~~\\ \hline
~~~~F~~~~&~~2.2566~~&~~1.7000~~&~~15.673~~&~~8.7698~~\\ \hline 
~~~~G~~~~&~~2.8672~~&~~1.8033~~&~~19.465~~&~~10.583~~\\\hline
	     \end{tabular}~~~~~~
\caption{For the case of $N=2$, the eigenvalues $l^2$ of the equation (\ref{eqpertug}) 
for $(m,n)=(1,1),(1,2)$ with the parameter $\alpha=0.1, \gamma=0.02, \mu=0.1$, $\beta=0.0,-0.15$.}
\label{tab:stabev_gravity}
\end{table}

\section{Summary and discussions}
The aim of this paper was twofold: first studying the effect of a brane cosmological 
constant on the six dimensional baby-Skyrme brane model and second studying the stability of the corresponding solution.
The brane cosmological constant was modeled by inflating four dimensional slices, which is relevant 
in the context of inflationary scenario within braneworlds and also at the
level of the classical equations: the inflating slices indeed model a four dimensional positive 
cosmological constant.

Various types of solutions exist when a bulk cosmological constant is supplemented in the model. However, the inclusion of an additional brane cosmological constant has the drastic effect of reducing the number of different types of solutions to essentially three, according to the sign of the bulk cosmological constant. The three solutions are characterized by specific geometries of the space sustaining the extradimensions (namely opened, flat or closed for a negative, null or positive bulk cosmological constant respectively).

The occurence of these three geometries seems to be indeed a generic feature since the same phenomenon occurs for the different model considered in \cite{Brihaye:2006pi}. This idea follows from the fact that these three peculiar solutions are indeed vacuum solutions, they are further relevant for the asymptotic form of the solutions in the presence of localized matter fields.

More interestingly, the different types of solutions available in the case of vanishing brane cosmological constant seems to be interconnected once the brane cosmological constant is turned on; the two types of solutions available in absence of the brane cosmological constant are recovered as limits of these two branches for specific values of the parameters. These branches exist up to a maximal value of the brane cosmological constant.

The second part of this paper deals with the stability of the inflating baby-Skyrme branes constructed in the first part. Although the unperturbed equations do not depend on the details of the brane, the stability analysis might be different for different four dimensional Einstein spaces.

We established the equations for the stability of the complete baby-Skyrme branes. 
Since these coupled equations are particularly involved, we examined in more details the stability issue in two particular simplified situations: the one with fixed gravitational backgrounds and perturbed baby-Skyrme field, the second with fixed baby Skyrme field and perturbed gravitational field.

Surprisingly, our results for the case of fixed gravitational background shows that bound states of the baby-Skyrme excitation exist only for fluctuations winding more than one time (say $N$ times, recall that there is a cylindrical symmetry in the extradimensions) in the extradimensions. For $N=2$, we found no tachyonic instability while tachyonic modes develop for higher values of the winding. The perturbation of the baby Skyrme field can be alternatively seen as a particle bound on the brane; we provide a rough estimation of the mass of such particles and it turns out that they should be of the order of $10$ TeV, i.e. accessible in ongoing experiments.

The perturbation in the gravitational sector admits bound states for $N\ge1$. We found a tachyonic instability for $N=1$ while the latter disappears for $N=2$. This suggest that the stability issue of the solution depends crucially on the winding number of the fluctuation and on the sector explored. We believe that the solutions to the coupled problem will confirm the pattern guessed by these preliminary results.

Let us finally mention some applications of our model: in the context of inflationary models, the particule spectrum provided by the stability analysis might have an interesting phenomenological interpretation. As an example, one can imagine that the particles emitted by the branes might have a particular signature on the CMB. Also, for unstable modes, one could imagine that the instability might turn the inflating brane to non inflating brane, thus providing a dynamical end to inflation. Let us note however that these ideas are quite speculative and would deserve deep further investigations.

\begin{acknowledgments}
Y. B. and T.D. thanks the FNRS for financial support. N.S. appreciates the kind hospitality of the University of Mons. 
\end{acknowledgments}

\appendix
\section{Numerical method}
First, we decompose the eigeneqauion (\ref{eqpertum}) into coupled first order equation
for $u(y),v(y):=\phi(y)$
\begin{eqnarray}
&&\frac{dv(y)}{dy}=u(y) \\
&&\frac{du(y)}{dy}=u(y)+\Bigl(p_1(y) \frac{u(y)}{v(y)}+p_2(y)+p_3(y)\Bigr)v(y)
\end{eqnarray}
where
\begin{eqnarray}
&&p_1(y):=\frac{y}{n^2\sin^2 f+L^2}\biggl[(N^2-3)n^2f'\sin 2f+\frac{L'}{L}(n^2\sin^2 f-L^2)
-4\frac{M'}{M}(n^2\sin^2 f+L^2)\biggr] \nonumber \\ 
&&p_2(y):=\frac{y^2}{n^2\sin^2 f+L^2}\biggl[n^2\frac{L'}{L}f'\sin 2f
+n^2\cos 2f\bigl((N^2-3)f'^2+N^2-1\bigr) \nonumber \\
&&\hspace{4cm}-n^2\sin 2f\Bigl(f''+4f'\frac{M'}{M}\Bigr)
-\mu L^2(N^2-1)\cos f\biggr] \\
&&p_3(y):=-\frac{m^2}{M^2}y^2 
\end{eqnarray}
Here we use a coordinate $r:=e^{y}$ to become finer the meshpoint at the vicinity of the origin. 
Essentially the method is similar to former described integration method in Sec.\ref{4a}. 
The main difference in this case is the matching procedure.
We evaluate the matching point $y_m$ by following condition:
\begin{eqnarray}
p_1(y_m)u(y_m)+(p_2(y_m)+p_3(y_m))v(y_m)=0
\end{eqnarray}
We solve the equation both from the origin and the infinity. In order to match $v(y)$ at 
$y=y_m$, we multiply a constant to the outer solution $v(y)$.
Instead of using the Wronski determinant (\ref{eqdet}), we introduce 
an arbitrary $\delta$-functional potential at an intermediate value $y_m$: 
\be
V_\delta(y):=-\frac{[v'(y_m)]^{y_m+0}_{y_m-0}}{v(y_m)}~\delta (y-y_m)
\ee
Because, if the $\delta-$functional potential exists, the eigenfunction is continuous at the 
matching point but its derivative is not. Therefore the correction in terms of the first order perturbation 
\be
\Delta E= \int v^*(y)V_\delta(y)v(y)=-[v'(y_m)]^{y_m+0}_{y_m-0}v(y_m)
\ee
efficiently improves the eigenvalue, i.e., the eigenfunction. 
If the analysis reaches the correct eigenfunction, it no longer has discontinuity 
at all and the computation is successfully terminated.


\begin{thebibliography}{qq}

%\cite{ArkaniHamed:1998rs}
\bibitem{ArkaniHamed:1998rs}
  N.~Arkani-Hamed, S.~Dimopoulos and G.~R.~Dvali,
  %``The hierarchy problem and new dimensions at a millimeter,''
  Phys.\ Lett.\  B {\bf 429}, 263 (1998)
  [arXiv:hep-ph/9803315].
  %%CITATION = PHLTA,B429,263;%%

%\cite{ArkaniHamed:1998nn}
\bibitem{ArkaniHamed:1998nn}
  N.~Arkani-Hamed, S.~Dimopoulos and G.~R.~Dvali,
  %``Phenomenology, astrophysics and cosmology of theories with  sub-millimeter
  %dimensions and TeV scale quantum gravity,''
  Phys.\ Rev.\  D {\bf 59}, 086004 (1999)
  [arXiv:hep-ph/9807344].
  %%CITATION = PHRVA,D59,086004;%%

%\cite{Randall:1999ee}
\bibitem{Randall:1999ee}
  L.~Randall and R.~Sundrum,
  %``A large mass hierarchy from a small extra dimension,''
  Phys.\ Rev.\ Lett.\  {\bf 83}, 3370 (1999)
  [arXiv:hep-ph/9905221].
  %%CITATION = PRLTA,83,3370;%%

%\cite{Randall:1999vf}
\bibitem{Randall:1999vf}
  L.~Randall and R.~Sundrum,
  %``An alternative to compactification,''
  Phys.\ Rev.\ Lett.\  {\bf 83}, 4690 (1999)
  [arXiv:hep-th/9906064].
  %%CITATION = PRLTA,83,4690;%%

%\cite{Cohen:1999ia}
\bibitem{Cohen:1999ia}
  A.~G.~Cohen and D.~B.~Kaplan,
  %``Solving the hierarchy problem with noncompact extra dimensions,''
  Phys.\ Lett.\  B {\bf 470}, 52 (1999)
  [arXiv:hep-th/9910132].
  %%CITATION = PHLTA,B470,52;%%

%\cite{Gregory:1999gv}
\bibitem{Gregory:1999gv}
  R.~Gregory,
  %``Nonsingular global string compactifications,''
  Phys.\ Rev.\ Lett.\  {\bf 84}, 2564 (2000)
  [arXiv:hep-th/9911015].
  %%CITATION = PRLTA,84,2564;%%

%\cite{Gherghetta:2000qi}
\bibitem{Gherghetta:2000qi}
  T.~Gherghetta and M.~E.~Shaposhnikov,
  %``Localizing gravity on a string-like defect in six dimensions,''
  Phys.\ Rev.\ Lett.\  {\bf 85}, 240 (2000)
  [arXiv:hep-th/0004014].
  %%CITATION = PRLTA,85,240;%%

%\cite{Giovannini:2001hh}
\bibitem{Giovannini:2001hh}
  M.~Giovannini, H.~Meyer and M.~E.~Shaposhnikov,
  %``Warped compactification on Abelian vortex in six dimensions,''
  Nucl.\ Phys.\  B {\bf 619}, 615 (2001)
  [arXiv:hep-th/0104118].
  %%CITATION = NUPHA,B619,615;%%

%\cite{Peter:2003zg}
\bibitem{Peter:2003zg}
  C.~Ringeval, P.~Peter and J.~P.~Uzan,
  %``Stability of six-dimensional hyperstring braneworlds,''
  Phys.\ Rev.\  D {\bf 71}, 104018 (2005)
  [arXiv:hep-th/0301172].
  %%CITATION = PHRVA,D71,104018;%%

\bibitem{Brihaye:2003ur}
Y.~Brihaye and B.~Hartmann,
%     title     = "{Born-Infeld strings in brane worlds}",
Nucl.\ Phys. {\bf B691}, 79-90 (2004)
[arXiv:hep-th/0311121]
 



%\cite{Roessl:2002rv}
\bibitem{Roessl:2002rv}
  E.~Roessl and M.~Shaposhnikov,
  %``Localizing gravity on a 't Hooft-Polyakov monopole in seven dimensions,''
  Phys.\ Rev.\  D {\bf 66}, 084008 (2002)
  [arXiv:hep-th/0205320].
  %%CITATION = PHRVA,D66,084008;%%

%\cite{Skyrme:1961vq}
\bibitem{Skyrme:1961vq}
  T.~H.~R.~Skyrme,
  %``A Nonlinear field theory,''
  Proc.\ Roy.\ Soc.\ Lond.\  A {\bf 260} (1961) 127.
  %%CITATION = PRSLA,A260,127;%%

%\cite{Faddeev:1996zj}
\bibitem{Faddeev:1996zj}
  L.~D.~Faddeev and A.~J.~Niemi,
  %``Knots and particles,''
  Nature {\bf 387}, 58 (1997)
  [arXiv:hep-th/9610193].
  %%CITATION = NATUA,387,58;%%

%\cite{Piette:1994jt}
\bibitem{Piette:1994jt}
  B.~M.~A.~Piette, W.~J.~Zakrzewski, H.~J.~W.~Mueller-Kirsten and D.~H.~Tchrakian,
  %``A Modified Mottola-Wipf model with sphaleron and instanton fields,''
  Phys.\ Lett.\  B {\bf 320}, 294 (1994).
  %%CITATION = PHLTA,B320,294;%%

%\cite{Piette:1994ug}
\bibitem{Piette:1994ug}
  B.~M.~A.~Piette, B.~J.~Schroers and W.~J.~Zakrzewski,
  %``Multi - Solitons In A Two-Dimensional Skyrme Model,''
  Z.\ Phys.\  C {\bf 65}, 165 (1995)
  [arXiv:hep-th/9406160].
  %%CITATION = ZEPYA,C65,165;%%

%\cite{Kudryavtsev:1996er}
\bibitem{Kudryavtsev:1996er}
  A.~E.~Kudryavtsev, B.~Piette and W.~J.~Zakrzewski,
  %``Mesons, baryons and waves in the baby Skyrmion model,''
  Eur.\ Phys.\ J.\  C {\bf 1}, 333 (1998)
  [arXiv:hep-th/9611217].
  %%CITATION = EPHJA,C1,333;%%

%\cite{Kodama:2008xm}
\bibitem{Kodama:2008xm}
  Y.~Kodama, K.~Kokubu and N.~Sawado,
  %``Localization of massive fermions on the baby-skyrmion branes in 6
  %dimensions,''
  Phys.\ Rev.\  D {\bf 79}, 065024 (2009)
  [arXiv:0812.2638 [hep-th]].
  %%CITATION = PHRVA,D79,065024;%%

%\cite{Kodama:2007kr}
\bibitem{Kodama:2007kr}
  Y.~Kodama, K.~Kokubu and N.~Sawado,
  %``Localizing gravity on Maxwell gauged CP1 model in six dimensions,''
  Phys.\ Rev.\  D {\bf 78}, 045001 (2008)
  [arXiv:0712.0700 [hep-th]].
  %%CITATION = PHRVA,D78,045001;%%

\bibitem{chovil}
I.~Cho and A., Vilenkin
%, A. Gravity of superheavy higher dimensional global defects 
Phys. Rev. D, 2003, 68, 025013


\bibitem{Brihaye:2006cs}
Y.~Brihaye and T.~Delsate,
%Eprint = {gr-qc/0605039},
Class.\ Quant.\ Grav. {\bf 24}, 1279-1292 (2007)
[arXiv:gr-qc/0605039]
%Inflating brane inside hyper-spherically symmetric defects


%\cite{Brihaye:2006pi}
\bibitem{Brihaye:2006pi}
  Y.~Brihaye, T.~Delsate and B.~Hartmann,
  %``Inflating branes inside abelian strings,''
  Phys.\ Rev.\  D {\bf 74}, 044015 (2006)
  [arXiv:hep-th/0602172].
  %%CITATION = PHRVA,D74,044015;%%

%\cite{Giovannini:2001xg}
\bibitem{Giovannini:2001xg}
  M.~Giovannini,
  %``Localization of metric fluctuations on scalar branes,''
  Phys.\ Rev.\  D {\bf 65}, 064008 (2002)
  [arXiv:hep-th/0106131].
  %%CITATION = PHRVA,D65,064008;%%

%\cite{Giovannini:2002jfa}
\bibitem{Giovannini:2002jfa}
  M.~Giovannini,
  %``Vector field localization and negative tension branes,''
  Phys.\ Rev.\  D {\bf 65}, 124019 (2002)
  [arXiv:hep-th/0204235].
  %%CITATION = PHRVA,D65,124019;%%

%\cite{Giovannini:2006ye}
\bibitem{Giovannini:2006ye}
  M.~Giovannini,
  %``Gravitating multidefects from higher dimensions,''
  Phys.\ Rev.\  D {\bf 75}, 064023 (2007)
  [arXiv:hep-th/0612104].
  %%CITATION = PHRVA,D75,064023;%%

%\cite{Giovannini:2002sb}
\bibitem{Giovannini:2002sb}
  M.~Giovannini,
  %``Gauge field localization on Abelian vortices in six dimensions,''
  Phys.\ Rev.\  D {\bf 66}, 044016 (2002)
  [arXiv:hep-th/0205139].
  %%CITATION = PHRVA,D66,044016;%%

%\cite{Giovannini:2002mk}
\bibitem{Giovannini:2002mk}
  M.~Giovannini, J.~V.~Le Be and S.~Riederer,
  %``Zero modes of six-dimensional Abelian vortices,''
  Class.\ Quant.\ Grav.\  {\bf 19}, 3357 (2002)
  [arXiv:hep-th/0205222].
  %%CITATION = CQGRD,19,3357;%%

%\cite{RandjbarDaemi:2002pq}
\bibitem{RandjbarDaemi:2002pq}
  S.~Randjbar-Daemi and M.~Shaposhnikov,
  %``A formalism to analyze the spectrum of brane world scenarios,''
  Nucl.\ Phys.\  B {\bf 645}, 188 (2002)
  [arXiv:hep-th/0206016].
  %%CITATION = NUPHA,B645,188;%%

%\cite{Chen:2000at}
\bibitem{Chen:2000at}
  J.~W.~Chen, M.~A.~Luty and E.~Ponton,
  %``A critical cosmological constant from millimeter extra dimensions,''
  JHEP {\bf 0009}, 012 (2000)
  [arXiv:hep-th/0003067].
  %%CITATION = JHEPA,0009,012;%%

%\cite{Kehagias:2000au}
\bibitem{Kehagias:2000au}
  A.~Kehagias and K.~Tamvakis,
  %``Localized gravitons, gauge bosons and chiral fermions in smooth spaces
  %generated by a bounce,''
  Phys.\ Lett.\  B {\bf 504}, 38 (2001)
  [arXiv:hep-th/0010112].
  %%CITATION = PHLTA,B504,38;%%

\bibitem{p-c-method}
``Numerical Recipes in C: The Art of Scientific Computing'',
Saul A.Teukolsky,William H.Press,William T,Vetterling
(Cambridge University Press, UK, 1988).
http://www.nr.com/oldverswitcher.html; 
{\it see also}, N.Watanabe, http://www-cms.phys.s.u-tokyo.ac.jp/$\sim$naoki/.

%\cite{Cooper:1994eh}
\bibitem{Cooper:1994eh}
  F.~Cooper, A.~Khare and U.~Sukhatme,
  %``Supersymmetry and quantum mechanics,''
  Phys.\ Rept.\  {\bf 251}, 267 (1995)
  [arXiv:hep-th/9405029].
  %%CITATION = PRPLC,251,267;%%

\bibitem{Eslami:2000tj}
P.~Eslami, W.~J.~Zakrzewski and M.~Sarbishaei,
Nonlinearity {\bf 13} 1867-81 (2000)
[arXiv:hep-th/0001153]
%Title = {{Baby Skyrme models for a class of potentials}},

\bibitem{colsys}
J.~ C.~U.~Ascher and R.~D.~Russell
% A collocation solver for mixed order systems of boundary value problems 
Math.\ of\ Comp. {\bf 33}, 659 (1979)



\end{thebibliography}
\end{document}